\RequirePackage{fix-cm}
\RequirePackage{amsmath}
\documentclass{svjour3}                     
\smartqed  
\usepackage{mathptmx}      
\usepackage{latexsym}
\usepackage{amsfonts}
\usepackage{amssymb}
\usepackage{amsmath}
\usepackage{graphicx}
\usepackage{tikz}
\usepackage{tikz-cd}
\usepackage{graphicx}
\usepackage{scalefnt}
\usepackage{algorithmic}
\usepackage{tikz-qtree}
\usetikzlibrary{patterns,matrix,arrows,decorations.pathmorphing}
\usetikzlibrary{shapes,positioning,chains,fit,calc}
\usetikzlibrary{decorations.markings}
\tikzstyle arrowstyle=[scale=1]
\tikzstyle directed=[postaction={decorate,decoration={markings,
		mark=at position .65 with {\arrow[arrowstyle]{stealth}}}}]
\tikzstyle reverse directed=[postaction={decorate,decoration={markings,
		mark=at position .65 with {\arrowreversed[arrowstyle]{stealth};}}}]

\usepackage{pgfplots}
\pgfplotsset{compat=1.15}
\usepackage{multirow}
\usepackage{epigraph,varwidth}
\usepackage{calc}
\usepackage{url}
\usepackage{mathtools, cuted}
\usepackage[section]{placeins}
\usepackage{algorithm}
\usepackage{subfigure}
\usepackage{breqn}
\let\originalepigraph\epigraph
\renewcommand\epigraph[2]%
{\setlength{\epigraphwidth}{\widthof{\mytextformat#1}}\originalepigraph{#1}{#2}}

\usepackage[colorlinks=true,citecolor=blue,linkcolor=magenta]{hyperref}

\newcommand{\comp}[1]{\overline{#1}\hspace{0.05cm}}
\newcommand\boldblue[1]{\textcolor{blue}{\textbf{#1}}}

\begin{document}

\title{Leveraging Special-Purpose Hardware for Local Search Heuristics\thanks{This work was supported by Fujitsu Laboratories of America, Inc.}}


\titlerunning{Leveraging Special-Purpose Hardware}        

\author{Xiaoyuan Liu        \and
        Hayato Ushijima-Mwesigwa \and
        Avradip Mandal \and
        Sarvagya Upadhyay \and
        Ilya Safro \and
        Arnab Roy
}


\institute{Xiaoyuan Liu \at
              School of Computing, Clemson University
              \email{xiaoyu3@clemson.edu}
           \and
           Hayato Ushijima-Mwesigwa \at
              Fujitsu Laboratories of America, Inc.
              \email{hayato@fujitsu.com}
              \and
           Avradip Mandal \at
              Fujitsu Laboratories of America, Inc.
              \email{amandal@fujitsu.com}
              \and
           Sarvagya Upadhyay \at
              Fujitsu Laboratories of America, Inc.
              \email{supadhyay@fujitsu.com}
              \and
           Ilya Safro \at
              School of Computing, Clemson University
              \email{isafro@clemson.edu}
              \and
           Arnab Roy \at
              Fujitsu Laboratories of America, Inc.
              \email{aroy@fujitsu.com}
}

\date{Received: date / Accepted: date}

\maketitle

\begin{abstract}
As we approach the physical limits predicted by Moore's law, a variety of specialized hardware is emerging to tackle specialized tasks in different domains. Within combinatorial optimization, adiabatic quantum computers, complementary metal oxide semiconductor (CMOS) annealers, and optical parametric oscillators are few of the emerging specialized hardware technology aimed at solving optimization problems. In terms of mathematical framework, the Ising optimization model unifies all of these emerging special-purpose hardware. In other words, they are all designed to solve optimization problems expressed in the Ising model or equivalently as a quadratic unconstrained binary optimization (QUBO) model. Due to variety of constraints specific to each type of hardware, they usually suffer from a major challenge: the number of variables that the hardware can manage to solve is very limited. Given that large-scale practical  problems, including problems in operations research, combinatorial scientific computing, data science and network science require significantly more variables to model than these devices provide, we are likely to witness that cloud-based deployments of these devices will be available for parallel and shared access. Thus hybrid techniques in combination with both hardware and software must be developed to utilize these technologies. The local search meta-heuristics is one of the approaches to tackle large scale problems. However, a general optimization step within local search is not traditionally formulated in the Ising form. In this work, we propose a new meta-heuristic to model local search in the Ising form for the special-purpose hardware devices. As such, we demonstrate that our method takes the limitations of the Ising model and current hardware into account, utilizes a given hardware more efficiently compared to previous approaches, while also producing high quality solutions compared to other well-known meta-heuristics.

\end{abstract}

\section{Introduction}
\label{sec:introduction}

Within the field of combinatorial optimization, driven by the physical limitations arising as a corollary of Moore's law \cite{schaller1997moore}, various research groups and institutions have started to develop novel hardware specifically designed for combinatorial optimization. Examples of such special-purpose hardware include adiabatic quantum computers \cite{johnson2011quantum}, complementary metal oxide semiconductor (CMOS) annealers \cite{aramon2019physics,yamaoka201524,yoshimura2013spatial} and coherent Ising machines \cite{inagaki2016coherent,kielpinski2016information,mcmahon2016fully}. Gate based quantum computers can also be used to solve such optimization problems, but are able to perform other tasks as well. Although these emerging technologies exhibit novelty in terms of hardware, they are all unified by the mathematical framework of the Ising optimization model. In other words, they are all designed to solve optimization problems formulated in the Ising model or equivalently as a quadratic unconstrained binary optimization (QUBO) problem. According to \cite{glover2018tutorial,kochenberger2006unified}, these recent hardware advances enable the Ising or QUBO model to become a unifying framework for combinatorial optimization. They have recently been termed as \emph{Ising processing units} (IPUs) \cite{coffrin2019evaluating}.

The focus of industry practitioners on development of specialized hardware to solve QUBO stems from the many advantages that QUBO offers. QUBO formulation can efficiently and succinctly abstract away real-world problems in network science \cite{shaydulin2018community,shaydulin2019network,ushijima2017graph},  chemistry \cite{hernandez2017enhancing,hernandez2016novel,terry2019quantum}, finance \cite{rosenberg2016solving}, and machine learning \cite{crawford2016reinforcement,henderson2018leveraging,khoshaman2018quantum,levit2017free,negre2019detecting}. The abstraction allows a cleaner and simpler mathematical framework to study a variety of problems arising from different disciplines. Many NP-hard combinatorial optimization problems can also be easily and efficiently reformulated in QUBO  \cite{glover2018tutorial,lucas2014ising}. While some of them admit integer programming formulation that can be converted into QUBO using standard techniques, others are necessarily formulated as QUBO. Unsupervised learning techniques such as spectral clustering, statistical neural models \cite{schneidman2006}, social network analysis problems such as community detection can be naturally cast as QUBO. The model's ubiquity and ability to represent a wide range of scientific problems makes development of specialized hardware for solving it a fruitful academic as well as industrial endeavor.

Given that QUBO is NP-hard in general, many heuristics have been developed to produce good solutions for large instances in a reasonable amount of time. Examples include but are not limited to tabu search \cite{glover1998adaptive,wang2012path}, simulated annealing (SA) \cite{kirkpatrick1983optimization}, large neighborhood search \cite{hamze2004fields,selby2014efficient}, integer programming \cite{dash2013note,mcgeoch2013experimental,puget2018d}, adiabatic quantum computation (AQC) \cite{farhi2000quantum,kadowaki1998quantum}, and quantum Monte Carlo \cite{nightingale1998quantum}. Novel hardware are being developed based on these methods. For example, the D-Wave quantum annealer is based on AQC, and the Fujitsu's digital annealer is  based on SA with parallel tempering.

\paragraph{Challenges with local search, QUBO, and IPU:}

Due to their size or complexity status, many computationally hard problems in combinatorial optimization require using heuristics that either decompose a large-scale problem and solve many smaller problems in parallel or iteratively improve some feasible solution for sufficiently many steps. Such algorithms as Kernighan-Lin \cite{kernighan1970efficient}, Fiduccia-Mattheyses \cite{fiduccia1982linear}, 2-sum window minimization \cite{safro2006multilevel}, and max-flow min-cut refinement \cite{sanders2011engineering} are among many other relevant examples applied in existing solvers. All these can be formulated as versions of a local search (or improvement) strategy that gradually improves a solution. With the advent of IPUs, and anticipating a hybridization of IPU and HPC systems, we envision that such local search problems (whose complexity are often the same as the original problem) will be solved on IPU devices using QUBO formulation. Typically, the larger a local search sub-problem, the better it can affect the global solution. However, increasing the size of sub-problem comes with a price tag of its representation as a QUBO on an IPU.

Some of the major challenges of QUBO hardware include limited precision and the maximum number of variables the hardware can handle. For instance, D-Wave's 2000Q quantum annealer, even with up to 2000 qubits, can only handle arbitrary fully connected QUBO of maximum 64 binary variables due to the connectivity of the hardware architecture. Fujitsu's latest quantum inspired digital annealer can handle up to 8192 binary variables and up to 64 bits of precision. This brings into perspective that current optimization problems arising from real-world challenges can easily have millions of variables. Subsequently, this motivates the investigation of the algorithmic challenge of how to utilize these emerging technologies to efficiently solve large scale problems.

Another challenge lies inherently within the QUBO model. Many problems in practice are coupled with multiple constraints; however, QUBO are unconstrained by definition. The natural strategy to turn a constrained problem into an unconstrained one is to introduce quadratic penalty terms (corresponding to each constraint) to the objective function. These penalties are introduced such that (i) if the constraint is satisfied, their contribution to objective function is exactly zero; and (ii) if the constraint is not satisfied, their contribution to objective function is negative (positive) for maximization (minimization) formulation. This leads to the need to tune the coefficients of the penalty terms. A small penalty term would easily lead to a violation of the constraint while a very large term can lead to difficulties in  comparison of quality of feasible solutions, especially when bounded by the limited precision of the hardware. In theory, we can derive a lower bound on the coefficients to enforce the constraints as demonstrated in \cite{lucas2014ising}, and the principle is to make sure that the penalty from the violation of constraint is larger than any possible changes of the objective function value in the original constrained problem. However, in practice, the coefficients are often chosen smaller than the bound to achieve a better result. Another consequence of this is that not all solutions to the QUBO formed would be a feasible solution to the original problem. Thus for some problems, finding a feasible solution via the QUBO model may become difficult, yet finding a feasible solution in their original optimization problem may be trivial. For example, problems arising from the permutations of $n$ objects lead to the formulation of a QUBO of size $\mathcal{O}(n^2)$. This QUBO would have $2^{n^2}$ candidate solutions, however only $n!$ of them would be a candidate permutation. Since
\begin{equation*}
    \frac{n!}{2^{n^2}} \sim \sqrt{2 \pi n} \left(\frac{n}{2^ne}\right)^n < \left(\frac{n}{2^{n}}\right)^n \longrightarrow 0,
\end{equation*}
 as $n \to \infty$, where $\sim$ means that the two values are asymptotic, we see that a candidate solution of the QUBO generated uniformly at random is not a permutation with high probability. Thus a straight-forward QUBO formulation of a permutation does not yield a good utilization of IPUs.

\paragraph{Our contribution:}

Similar to previous methods \cite{shaydulin2019hybrid,shaydulin2018community,shaydulin2019network}, for problems larger than current hardware size, we advocate the use of a local search framework to create sequence of sub-problems that can be solved with hardware of limited size.  We name this framework as QUBO local search (QUBO-LS). In addition, we categorize QUBO-LS into two types: constrained QUBO local search (C-QUBO-LS) and unconstrained QUBO local search (U-QUBO-LS). In particular, our U-QUBO-LS  addresses the limitations of QUBO framework and demonstrates efficient use of the hardware by formulating the local search sub-problems into QUBOs that use less binary variables and more tailored to the original combinatorial optimization problems. Our U-QUBO-LS can be easily generalized to different combinatorial optimization problems, and we give models for local search technique for the traveling salesman problem (TSP), graph partitioning (GP), quadratic assignment problem (QAP) and minimum 2-sum problem (M2sP) on IPUs as examples.

The IPU experiments are carried out using Fujitsu's latest digital annealer. In order to show the differences in modeling methods, we compare our approach to previous methods on QAP and M2sP. We anticipate that in the near future, availability of these devices in combination with HPC will play an important role in breaking the barriers of existing solvers and this type of modeling will be broadly applicable.

\section{Background}
\label{sec:background}

\subsection{Ising Model}
\label{sec:background1}

The Ising model is a common mathematical abstraction which has been widely used in physics. In this class of graphical models, the nodes $\mathcal{N}$ represent discrete spin variables (i.e., $\sigma_i \in \{-1, 1\}, \forall i \in \mathcal{N}$), and the edges $\mathcal{E}$ represent the interactions of spin variables (i.e., $\sigma_i\sigma_j, \forall (i, j) \in \mathcal{E}$). For each node, a local field $h_i, \forall i\in\mathcal{N}$ is specified, and for each edge, an interaction strength $J_{ij}, \forall (i,j) \in\mathcal{E}$ is specified. The energy of a configuration $\sigma$ is given by the Hamiltonian function: \begin{eqnarray}\label{eq:hamiltonian}
H(\sigma) = \sum_{(i,j)\in\mathcal{E}} J_{ij}\sigma_i\sigma_j + \sum_{i\in\mathcal{N}} h_i\sigma_i.
\end{eqnarray}

The most common applications of the Ising model is to find the lowest possible energy of the model, namely, to find the configuration $\sigma$ that minimizes the Hamiltonian function (\ref{eq:hamiltonian}). Note that with a transformation of the variables, $\sigma_i = 2x_i - 1, i\in \mathcal{N}$, where $x_i \in \{0, 1\}$, an Ising optimization problem is equivalent to QUBO.

\subsection{Digital Annealer}
\label{sec:background2}

Fujitsu's Digital Annealer (DA) is a hardware accelerator for solving fully connected QUBO problems (i.e., the values of $J_{ij}$ are nonzero for all $i, j\in \mathcal{E}$ in Equation (\ref{eq:hamiltonian})). Internally the hardware runs a modified version of Metropolis-Hastings algorithm \cite{hastings1970monte,metropolis1953equation} for simulated annealing. The hardware utilizes massive parallelization and a novel sampling technique. The novel sampling technique speeds up the traditional Markov Chain Monte Carlo (MCMC) method by almost always moving to a new state instead of being stuck in a local minimum. As explained in \cite{aramon2019physics}, in the DA, each Monte Carlo step takes the same amount of time, regardless of accepting a variable flip or not. In addition, when accepting the flip, the computational complexity of updating the effective fields is constant regardless of the connectivity of the graph. DA also supports parallel tempering (replica exchange MCMC sampling) \cite{swendsen1986replica} which improves dynamic properties of the Monte Carlo method. In our experiments we used this mode, as it requires less parameter tuning and better for consistent benchmarking. We used the second generation of the DA also known as the \emph{Digital Annealing Unit} (DAU), which supports $8192$ binary variables with up to $64$ bits of precision for the individual entries of the weight matrix.

\subsection{Related Work}
\label{sec:background3}
With respect to solving problems larger than the current hardware can accommodate, a large number of work has focused on formulating the original combinatorial problem as a large QUBO and then using some decomposition technique to create  sub-QUBOs that can be individually solved directly on the hardware. The tool qbsolv \cite{booth2017partitioning} developed by D-Wave systems is one such example. For example, with the limitation that the D-Wave 2X and D-Wave 2000Q quantum annealers can solve fully dense QUBO of up to approximately 45 and 64 binary variables respectively, researchers at Volkswagen \cite{neukart2017traffic} solved a traffic flow optimization problem that modeled traffic from 418 cars that required 1254 binary variables. In \cite{negre2019detecting,ushijima2017graph}, problems in graph partitioning and community detection were solved for graphs larger than hardware size.

An alternative approach to solve problems larger than the hardware size is to first identify a subproblem, and then model this subproblem as a QUBO and solve with a given hardware. This type of approach is usually framed as large neighborhood search (LNS) meta-heuristic \cite{pisinger2010large}. It explores a complex and large neighborhood and makes it possible to find better solution in each iteration. For example, the authors in \cite{shaydulin2019hybrid,shaydulin2018community,shaydulin2019network,multilevel} generally took this approach for solving the graph partitioning and community detection problem on available quantum computing hardware. They solved problems on the D-Wave quantum annealer and the gate-model IBM quantum computer by creating and solving smaller QUBOs. Their approach is referred to as the quantum local search (QLS). However, a straight forward extension of QLS applied to a general combinatorial optimization problem does not take the limitations of the QUBO model into account. In this work, we consider modeling of sub-problems while taking limitations of the QUBO model into account. Moreover, we show that we can embed different local search heuristics into the QUBO model, thus making the QUBO-LS framework more generic and can be extended to other combinatorial optimization problems.

\section{QUBO Local Search}
\label{sec:modeling}

\begin{figure*}
\centering
\hfill
\subfigure[Before: $Cut = 6$]{\includegraphics[width=0.45\linewidth]{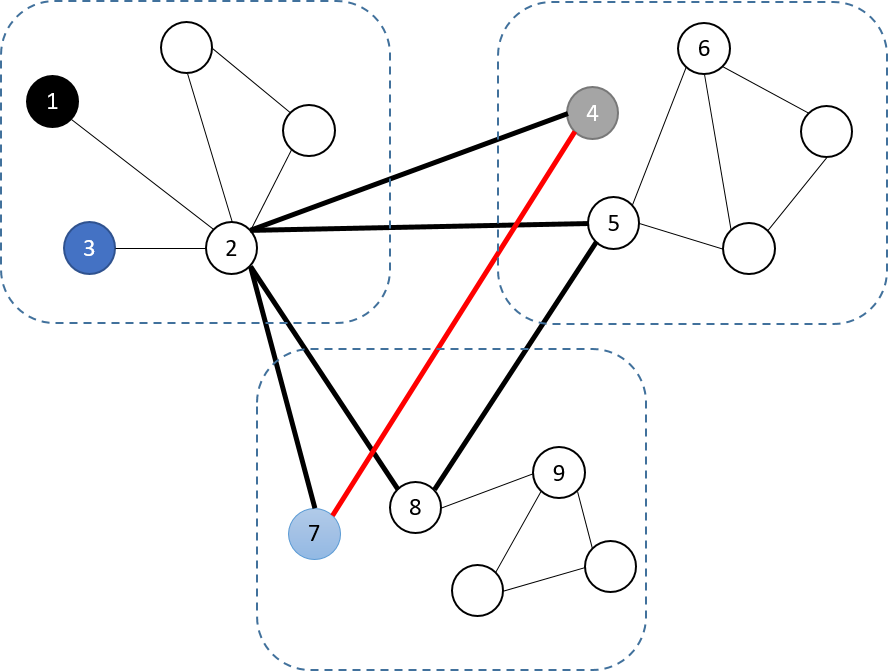}}
\hfill
\subfigure[After: $Cut = 5$]{\includegraphics[width=0.45\linewidth]{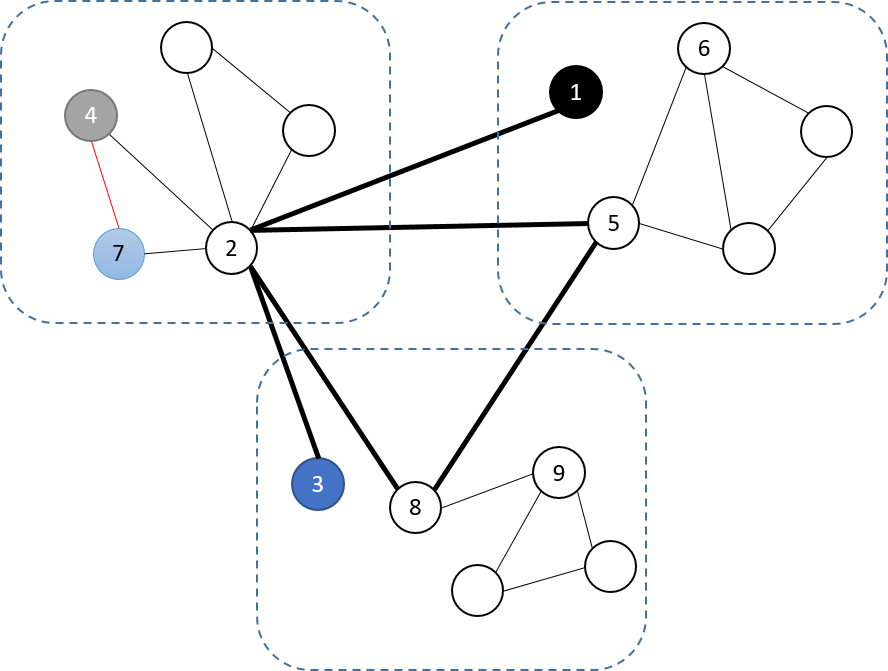}}
\hfill
\caption{Performing the swaps, 3 with 7, and 1 with 4 simultaneously results into a reduced objective value, where as the individual swaps would not improve the objective value.}
\label{fig:example}
\end{figure*}

\paragraph{Local search} is a class of metaheuristic methods for solving large scale combinatorial optimization problems. A local search algorithm moves from one feasible solution to another by applying local changes to the current solution. The use of local search in combinatorial optimization dates back to the 1950s when the first edge-exchange algorithms were introduced for the travelling salesman problem \cite{bock1958algorithm}. Since then it has been broadened with various levels of success in different problems. The scaling of Moore's law together with the use of sophisticated data structures has made local search algorithms the state-of-the-art for many problems. However, local search algorithms often get stuck in a local optima when there is no better solution to be found by only applying a single local change, but may be improved if two or more local changes are applied simultaneously. As an example, we use the balanced graph partitioning problem to illustrate this. The balanced graph partitioning problem aims to partition the vertices into equal parts such that the number of cut-edges is minimized, where a cut-edge is defined as an edge whose endpoints are in different parts.
The graph in Figure \ref{fig:example} has 15 nodes that are partitioned into 3 balanced parts as shown in Figure \ref{fig:example}(a). When we consider the pair of nodes $(1, 4)$ or $(3, 7)$ independently, we would not choose to swap them since the individual swaps will not reduce the cut. However, swapping both  pairs simultaneously improves the  partitioning to a cut of 5.

With the introduction of the special-purpose hardware for combinatorial optimization, in a post-Moore's law era, one of the interesting questions is whether we can develop hybrid techniques to fully utilize these technologies. However, since sub-problems within local search are not traditionally described with respect to the Ising model, it remains unclear how well established algorithms can take advantage of these technologies. In this section, we take steps to demonstrate the use of this technology. We do this by introducing different ways in which a sub-problem in existing algorithms can be modeled as a QUBO. We name this framework QUBO local search (QUBO-LS).

\subsection{Constrained and Unconstrained QUBO Local Search}
\label{sec:modeling1}

When formulating combinatorial optimization problems as binary optimization problems, additional constraints are often introduced. One of the most common types of constraints is  usually referred to as 1-hot constraints or 1-hot encoding. For $n$ binary variables $\{x_i\}_{i=1}^n$, a 1-hot encoding is simply the constraint of the form
\begin{equation}
    \sum_{i = 1}^n x_i = 1.
\end{equation}
For $n^2$ binary variables $\{x_{i,j}\}_{1\leq i,j\leq n}$, a 2-way 1-hot encoding are the constraints of the form
\begin{equation}
    \begin{aligned}
    \label{eq:1hot}
    \sum_{i=1}^n x_{i,j} = 1,  &\quad j = 1, \dots, n,\\
    \sum_{j=1}^n x_{i,j} = 1,  &\quad i = 1, \dots, n.
\end{aligned}
\end{equation}
The 2-way 1-hot encoding are particularly common because they model permutations which have common occurrence in combinatorial optimization problems.

If $\mathbf{x} = \{x_{i,j}\}_{1\leq i,j\leq n}$ and  $Q(\mathbf{x})$ is a quadratic function we want to minimize, subject to 2-way 1-hot constraints, it is well known, that for appropriate choice of positive constants $\lambda_i$ and $\lambda'_j$ 's, the above set of constraints can be encoded as a QUBO problem where the goal is to minimize
\begin{align}
    \label{eq:qubo-2way1hot}
    Q(\mathbf{x}) + \sum_{i=1}^n \lambda_i  \Big(\sum_{j=1}^n x_{i,j} - 1 \Big)^2 \\
    + \sum_{j=1}^n \lambda'_j  \Big(\sum_{i=1}^n x_{i,j} - 1 \Big)^2.\nonumber
\end{align}

If the above problem forms a sub-problem in a local search framework, the 1-hot encoding and 2-way 1-hot encoding would significantly reduce the feasible solution search space compared to if there are no such constraints. The formulations of the TSP, GP, QAP and M2sP all contain such constraints that reduce the feasible solution search space significantly. In the following sub-sections we use these problems as examples where sub-problems can be formulated as QUBOs that do not require any constraints and thus increases the search space the QUBO solver can search per iteration. For QAP, we also show how to extend QUBO-LS in which one can control the solution space searched per iteration of the QUBO solver. We name the type of QUBO-LS that contains penalty terms for constraints in the formulation as constrained QUBO local search (C-QUBO-LS). Similarly, we name the type of QUBO-LS that does not contain any penalty terms as unconstrained QUBO local search (U-QUBO-LS).

\subsection{Travelling Salesman Problem}
\label{sec:modeling2}

The travelling salesman problem is by far one of the most well known combinatorial optimization problems. Many algorithms, both heuristics and exact methods have been proposed. As such it is instructive to describe any new general approach  with respect to TSP for ease of exposition. Given a list of $n$ cities and the distances between each pair of cities, the goal of TSP is to find the shortest possible tour that visits each city exactly once and returns to the origin city. In a QUBO representation, it has a formulation with $n^2$ binary variables which makes solving current real size TSP instances directly on near term hardware unlikely. If we cannot directly solve a large TSP instance with hardware, the next best thing is to accelerate current TSP algorithms and heuristics. Current state-of-the-art methods consist of using local search with sophisticated data structures. From the hardware perspective, improving the speed or quality of local search moves seems like the best option to enhance current methods. The $k$-opt algorithm is one of the most popular heuristics for solving TSP. However, a straight-forward implementation of $k$-opt in C-QUBO-LS would require at least $k^2$ variables due to the 2-way 1-hot encoding of the sub-problem. Within a QUBO model, adding constraints reduces the number of feasible solutions searched per iteration. In this sub-section we give an alternative U-QUBO-LS model whose formulation does not require constraints thus can search at most up to $2^H$ feasible solutions per iteration, where $H$ is the hardware size.

\begin{figure}
\centering
\includegraphics[width=0.47\linewidth]{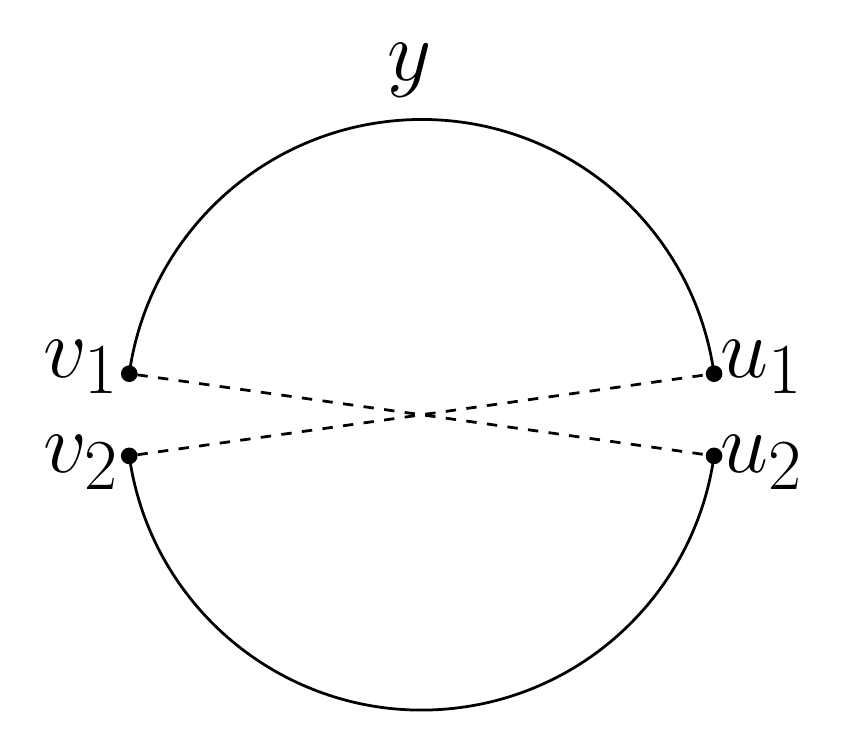}
\includegraphics[width=0.47\linewidth]{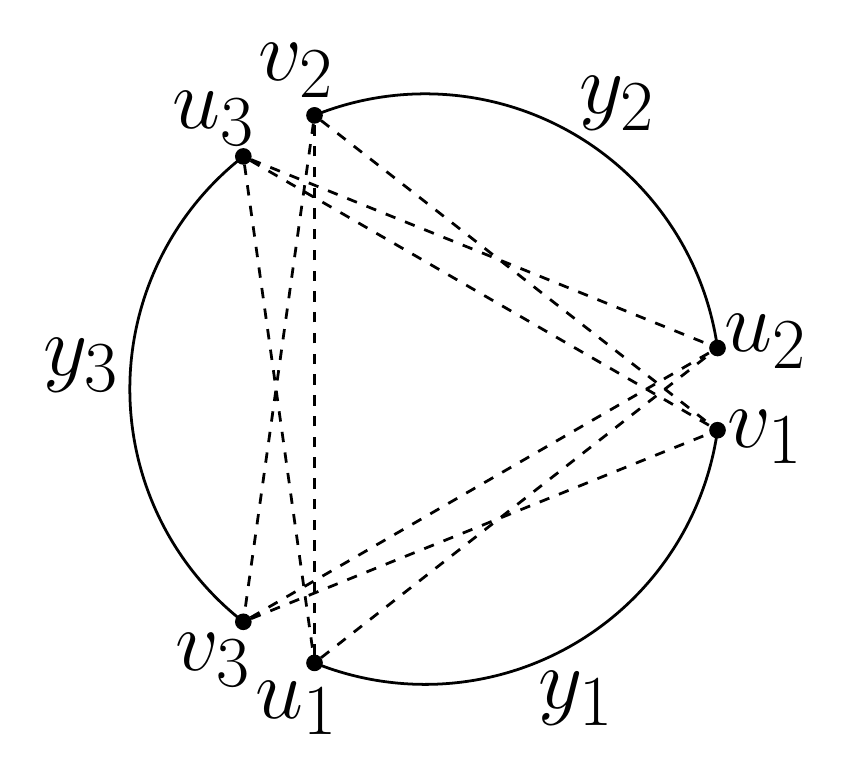}
\caption{2-Opt and 3-Opt sub-problems}
\label{fig:2opt}
\end{figure}

\paragraph{The $k$-opt algorithm} is a well-known local search heuristic for the TSP. A $k$-opt move consists of removing $k$ edges from a given tour and then reconnecting the $k$ segments to possibly get a shorter tour. In this sub-section, we use the TSP as an easy example to give a QUBO formulation of a sub-problem that does not contain 2-way 1-hot encoding although the original problem does. For ease of exposition, for a binary variable $y \in \{0,1\}$, we use the notation $\comp{y}$ such that
\begin{equation}
    \comp{y} = 1 - y.
\end{equation}
In the case of the 2-opt algorithm, a local search move is a decision of whether 2 edges in the tour should be replaced with 2 other edges. In an effort to motivate our modeling technique, we first model a move in 2-opt, and 3-opt as a QUBO. An equivalent and alternative way of stating a 2-opt move is as follows: remove two edges from the current tour, thus creating two disjoint paths which we shall refer to as segments. Then a 2-opt move is equivalent to decide whether or not to reverse one of these segments. A segment $(u_1, v_1)$ is \emph{reversed} if it appears in the new tour in the reversed order $(v_1, u_1)$. Therefore a 2-opt move represented as a U-QUBO-LS with one variable as

\begin{equation}
\begin{aligned}
\min_y \ & (w_{u_1v_2} + w_{u_2v_1})y + (w_{u_1u_2} + w_{v_1v_2})\comp{y}\\
\text{s.t. } & y \in \{0,1\},
\end{aligned}
\end{equation}
where $w_{u,v}$ denotes the distance between city $u$ and $v$. Figure \ref{fig:2opt} depicts a 2-opt and 3-opt move with respect to decision variables that constitute of reversing a segment or not.

\begin{figure}
	\centering
	\includegraphics[width=0.5\linewidth]{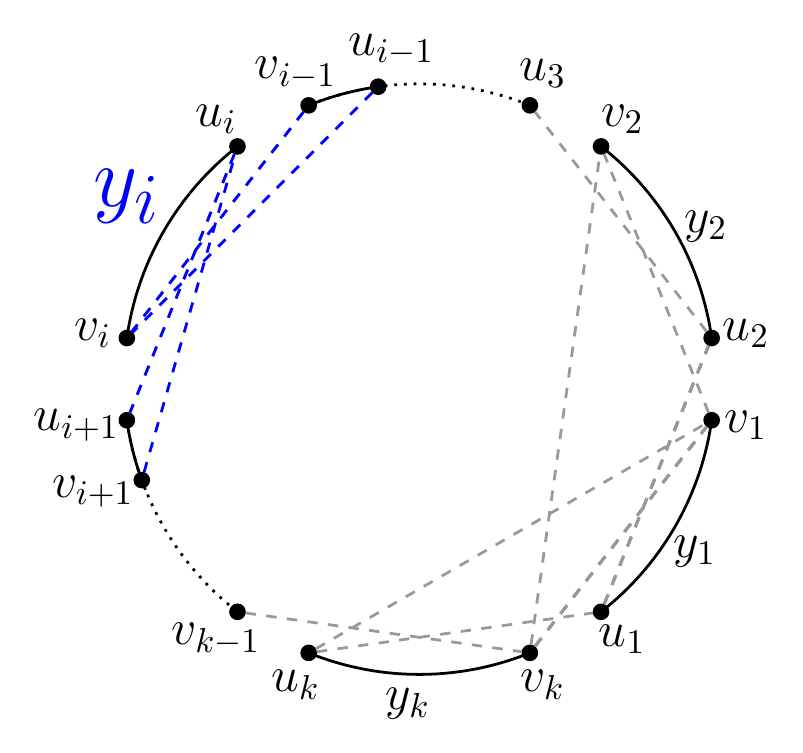}
	\caption{Sub-problem for TSP that reverses at most $k$ segments of the tour. This sub-problem is purely an unconstrained problem thus does not require modeling any constraints as a QUBO. As such, all $2^k$ solutions of the QUBO are also feasible solutions of the TSP}
	\label{fig:kopt}
\end{figure}

We can then extend this approach for larger $k$ and define a move as a $k$-reversal as shown in Figure \ref{fig:kopt}. At each iteration, a $k$-reversal would decide whether or not to reverse up to $k$ segments of the tour.
In general, since $y_i$ is 1 if segment $i$ is reversed and $\comp{y_i}$ is 1 if it is \emph{not} reversed, then product $y_i\comp{y_j}$ is 1 if and only if segment $i$ is reversed and segment $j$ is not. Thus, the quadratic terms can only be in the form of either $\comp{y_i}\comp{y_{j}}, \comp{y_i}y_{j},y_{i}\comp{y_j}$ or $y_iy_{j}$. Therefore, let
\begin{equation}
\begin{aligned}
    q(y_i, y_j) &= w_{v_iu_{j}}\comp{y_i} \comp{y_{j}} + w_{u_iu_{j}}y_i\comp{y_{j}}\\
&+ w_{v_iv_{j}}\comp{y_i}y_{j} +w_{u_iv_{j}}y_iy_{j},
\end{aligned}
\end{equation}
then $q(y_i, y_{i+1})$ and $q(y_{i-1}, y_i)$, contain all the quadratic terms in a $k$-reversal move with respect to $y_i$, for $i = 2, \dots, k-1$. Therefore, a $k$-reversal move represented as a U-QUBO-LS with $k$ variables as
\begin{equation}
\begin{aligned}
\min \  &q(y_k, y_{1}) + \sum_{i=1}^{k-1} q(y_i, y_{i+1}),\\
\text{s.t. } \quad &y_i \in \{0, 1\} \quad  i = 1, \dots, k.
\end{aligned}
\end{equation}
In particular, all 3-opt moves are also 3-reversal moves.
The main advantage of this model is that it does not model any constraints thus any  solution of the QUBO is a feasible solution of the TSP. In other words, at most $2^k$ feasible solutions are considered at each iteration of the local search. The major drawback of this approach is that it is not equivalent to a $k$-opt move. In particular, there are $(k-1)!2^{k-1}$ possible ways to reconnect $k$ segments. However, out of all these moves, $k$-reversal only considers $2^k$ of them. With this in mind, we believe that this model gives the reader a more intuitive understanding on how to model sub-problems of a given problem. In the next subsection, we discuss the Kernighan-Lin algorithm which is based on swapping nodes between parts. These moves can also be modeled as a U-QUBO-LS.

\subsection{Graph Partitioning Problem}
\label{sec:modeling3}
The graph partitioning problem (GP) is another well known combinatorial optimization problem with many applications \cite{bulucc2016recent}. Formally, let $G = (V, E)$ be an undirected graph of vertices $V$ and edges $E$. Let $|V|$ denote the number of vertices of the graph, and $w_{ij} \geq 0$ be the weight of the edge between nodes $i$ and $j$. For a fixed integer $K$, the $K$-way GP is to find a partition $\Pi = (\Pi_1, \cdots, \Pi_K)$ of the vertices $V$ into $K$ equal parts (here we discuss the perfectly balanced version of GP) such that the number of cut-edges is minimized, where a cut-edge is defined as an edge whose endpoints are in different partitions. The QUBO formulation for GP \cite{ushijima2017graph} is given as follows:
\begin{eqnarray*}
	\min_{\Pi} & & A\sum_{i=1}^{|V|}\left[\left(\sum_{\ell=1}^{K} x_{i\ell}\right) - 1\right]^2 + B\sum_{\ell=1}^K \left[\left(\sum_{i=1}^{|V|} x_{i\ell}\right) - \frac{|V|}{K}\right]^2 \\
	&+& \sum_{(i,j)\in E}\sum_{\ell=1}^Kw_{ij}(x_{i\ell}-x_{j\ell})^2,
\end{eqnarray*}
where $A, B > 0$ are constants to penalize the violation of constraints. For a feasible solution, the binary variables $x_{i\ell}$ are interpreted as follows: \begin{eqnarray*}x_{i\ell} = \begin{cases}
		1, & \mbox{if vertex } i \text{ is in partition } \ell\\
		0, & \mbox{otherwise}.
	\end{cases}
\end{eqnarray*}

\paragraph{Kernighan-Lin (KL) algorithm} is a very popular algorithm for graph partitioning dating back to the seminal paper \cite{kernighan1970efficient} and used in a variety of multilevel solvers for graphs and hypergraphs  \cite{boman2012zoltan,karypis1998fast,safro2015advanced,shaydulin2019relaxation}. The KL algorithm is an iterative algorithm whose goal is to reduce the number of cut edges between two parts. The main concepts used in the algorithm can be described as follows.
Let $|V| = 2n$, $V_1, V_2, \subset V$ such that $|V_1| = n = |V_2|$ and $V_1 \cap V_2 =\text{\O}$. For $u \in V_1$ define
\begin{equation}
\begin{aligned}
  E_u = \sum_{v \in V_2} w_{uv}; \quad & I_u  = \sum_{v \in V_1} w_{uv},
\end{aligned}
\end{equation}
as the \emph{External} and \emph{Internal} degree of node $u \in V_1$ respectively. Let
\begin{equation}
    D_u = E_u - I_u,
\end{equation}
be the cut reduction of moving node $u \in V_1$ to  $V_2$. We refer to this as the $D$-value of $u$. Then the cut reduction from swapping $u$ and $v$ is given by
\begin{equation}
    g_{uv} = D_u + D_v - 2 w_{uv}.
\end{equation}
This is usually referred to as the \emph{gain} of swapping.  The KL algorithm attempts to find an optimal series of swapping operations between elements  in $V_1$ and $V_2$ which maximizes the gain of swapping and then executes the operations. We can model such local search process using a U-QUBO-LS formulation. Let $s: V_1 \to V_2$ be the one-to-one function that identifies a node in $V_2$ that will potentially be swapped given a node in $V_1$. Then let
\begin{equation}
    M = \big\{\{u, s(u)\} | u \in V_1 \big\}.
\end{equation}
For every $\{u,v\} \in M$, define the binary variable $y_{uv}$ such that
\begin{equation}
    y_{uv} = \begin{cases}
    1, & \text{if $u$ swaps with $v$},\\
    0, & \text{otherwise.}
    \end{cases}
\end{equation}
Since the function $s$ is a one-to-one function, we simplify the notation and refer to the variable $y_{uv}$ simply as $y_u$ or $y_v$. In other words,
\begin{equation}
    y_{uv} = y_{vu}= y_u = y_v.
\end{equation}
Thus $y_v$ simply represents the variable associated to moving node $v$ for any $v \in V$. Then we can write the external degree of a node $u \in V_1$ as
\begin{equation}
\begin{aligned}
    E_{u} &= \sum_{v \in V_2} w_{uv}(\comp{y_u}\comp{y_v} + y_u y_v)\\
    &+ \sum_{v \in V_1} w_{uv}(y_u\comp{y_v} + \comp{y_u} y_v).
\end{aligned}
\end{equation}

Since the sum of the external degree for every node in $V$ is in fact equal twice the cut, we thus have an optimal move as a U-QUBO-LS:
\begin{equation}
\begin{aligned}
    \min \ & \sum_{u \in V} E_u, \\
    & y_u \in \{0,1\} \quad u \in V.
\end{aligned}
\end{equation}
We now generalize this to give an optimal move in a $K$-way partitioning.

Let $P(u) \in \{1,\dots, K\}$ be the index of partition that node $u$ currently belongs to, i.e., $P(u) = \ell$ if $u \in V_{\ell}$. Let $M$ be the set of non-intersecting pairs of nodes being considered for the decision to be swapped, such that if $\{u_1, u_2\}, \{u_3, u_4\} \in M$, then  $u_i \neq u_j$ for $i\neq j$ and if $\{u,v\} \in M$ then $P(u) \neq P(v)$, i.e, they are distinct and nodes in each pair belong to different parts. We define this property of $M$ as \emph{pairwise disjoint}.  Similar to the bisection case, define $y_{uv} = y_{vu}= y_u = y_v$ as the binary variable that is 1 if and only if node $u$ swaps parts with $v$.
Define the \emph{community} of $u\in V$ as
\begin{equation}
    N(u) := \{ v \in V | P(u) = P(v) \},
\end{equation}
and
\begin{equation}
    \comp{N}(u) := \{ v \in V |  \{w, v\} \in M,   w \in N(u)\},
\end{equation}
then since each node is restricted to moving to only one other part, we can describe this move as a QUBO similar to the one in the graph bisection problem.

The external degree of a node $u\in V$ can be defined in terms of the disjoint sets $N(u)$ and $\comp{N}(u)$

\begin{equation}
    \begin{aligned}
 E_{u} &= \sum_{j \in N(u)} w_{uj}(y_{u}\comp{y_{j}} + \comp{y_u}y_j) \\
     &+ \sum_{j \in \comp{N}(u)} w_{uj}(y_{u}y_{j} + \comp{y_u}\comp{y_j}) \\
   & + \sum_{j \in V \setminus N(u) \cup \comp{N}(u)} w_{uj}.
    \end{aligned}
\end{equation}

Then for a partition given by $V_1, \dots, V_K$, the cut is given by
\begin{equation}
   \frac{1}{2} \sum_{u \in V} E_u,
\end{equation}
where the $\frac{1}{2}$ is added to include the double counting of each edge in the cut but can be ignored for optimization purposes. Therefore an optimal move of swaps would be the U-QUBO-LS
\begin{equation}
\begin{aligned}
    \min \ & \    \sum_{j=1}^{K}\sum_{u \in V_j} E_u, \\
    & y_u \in \{0,1\} \quad u \in V,
\end{aligned}
\end{equation}
or simply as
\begin{equation}
\begin{aligned}
    \min \ & \    \sum_{ \{u, v \} \in M} E_u + E_v,\\
    & y_{uv} \in \{0,1\} \quad \{u,v\} \in M,
\end{aligned}
\label{eq:kl_qubo}
\end{equation}
where the formulation given in (\ref{eq:kl_qubo}) is independent from $K$. Note that if there exists a node $u \in V$ such that $u \notin m$ for any $m \in M$, then by definition, $y_u =0$, thus not every node needs to be considered to be move at each iteration of the algorithm.

\subsection{Quadratic Assignment Problem}
\label{sec:modeling4}

Quadratic assignment problem (QAP) is one of the fundamental combinatorial optimization problems that generalizes many other famous problems including TSP and GP.

Consider a set of facilities $\mathcal{F} = \{1, \dots, n\}$ and a set of locations  $\mathcal{L} =  \{1, \dots, n\}$. We define flow weight $w_{ij}$ and distance $d_{kl}$ for all pairs of facilities ($i$ and $j$) and locations ($k$ and $l$), respectively. The quadratic assignment problem (QAP) is to assign all facilities to different locations with the goal of minimizing the allocation cost, taking the costs as the sum of all possible distance-flow products. The QAP can be formulated as the following QUBO: \begin{eqnarray}\label{eq:quboform}
\min & & \sum_{i=1}^n\sum_{j=1}^n\sum_{k=1}^n\sum_{\ell=1}^n w_{ij}d_{k\ell}x_{ik}x_{j\ell} \\
\notag & & + A\sum_{k=1}^{n}(\sum_{i=1}^{n}x_{ik} -1)^2 \\
\notag & & + B\sum_{i=1}^{n}(\sum_{k=1}^{n}x_{ik} -1)^2, \\
\notag \text{s.t.} & & x_{ik} \in \{0, 1\}, i, k = 1, 2, \cdots, n,
\end{eqnarray}
where $A, B > 0$ are constants to penalize the violation of constraints.

For a feasible solution, the binary variables $x_{ik}$ can be interpreted as follows: \begin{eqnarray*}
	x_{ik} = \begin{cases}
		1, & \mbox{if facility $i$ is assigned to location $k$,}  \\
		0, & \mbox{otherwise}.
	\end{cases}
\end{eqnarray*}
Note that this is a classical example of problem with 2-way 1-hot constraints, and each feasible solution is an encoding of a permutation.

\paragraph{Local search algorithm} As summarized in \cite{burkard1998quadratic}, much research has been devoted to the development of local search heuristics to provide good quality solutions of QAP in a reasonable time. All such algorithms start with an initial permutation (assignment) and iteratively refine the current permutation. The most frequently used neighborhoods for QAPs are the pair-exchange neighborhood and the cyclic triple-exchange neighborhood.

A straightforward implementation of quantum local search (QLS)  \cite{shaydulin2019hybrid,shaydulin2018community,shaydulin2019network} would be: given an initial permutation of QAP, in each iteration, select a subset of the current permutation, formulate a QUBO that searches through possible configuration inside the subset. The QUBO formulated in each iteration will have $k^2$ binary variables, where $k$ is the size of the subset selected. This formulation contains the 2-way 1-hot encoding penalty terms, and therefore is a C-QUBO-LS.

\begin{figure*}[!htbp]
	\centering
	\includegraphics{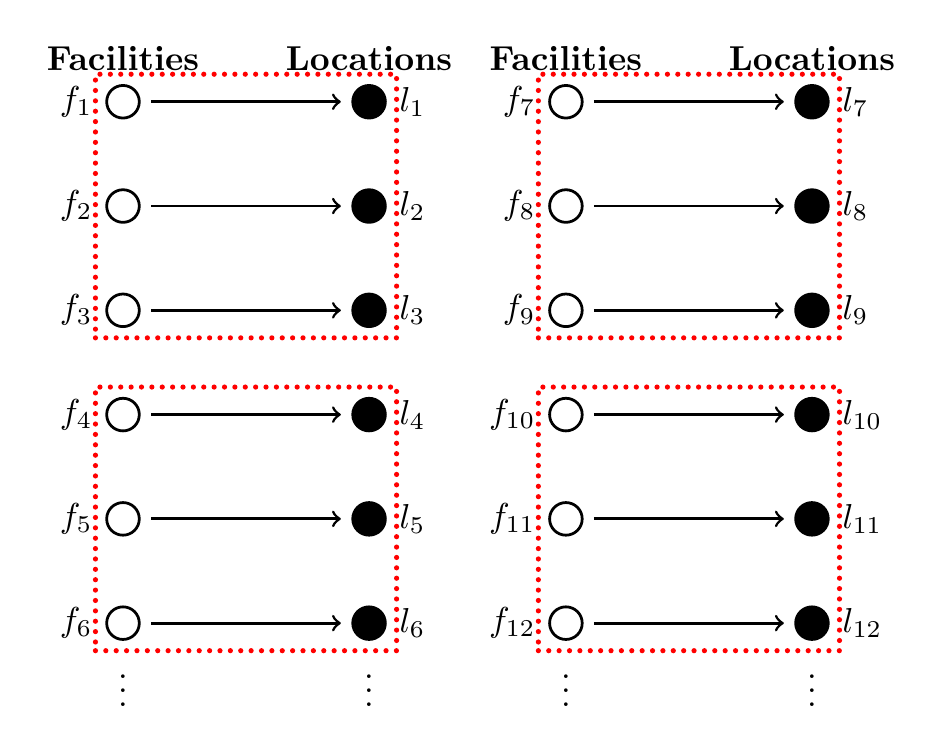}
	\caption{Multiple Local Search Simultaneously with $k=3$}
	\label{fig:multi_ls_k3}
\end{figure*}

We first extend QLS by choosing multiple subsets that are pairwise disjoint. Given an initial permutation of QAP, in each iteration, we first select $m$ pairwise disjoint subsets of size $k$ from the permutation (for the sake of simplicity, we assume that the subsets are equally sized, but in general they can be different in sizes). Next, we try to refine the current permutation by finding the optimal allocation of each subset we select simultaneously, and we achieve this by formulating it into a QUBO and solve the QUBO. To make it more clear, we use Figure \ref{fig:multi_ls_k3} to further explain. There are 4 disjoint subsets, each with size 3, thus $m=4, k=3$, we allow facility $f_1, f_2, f_3$ to relocate to location $l_1, l_2, l_3$, facility $f_4, f_5, f_6$ to relocate to location $l_4, l_5, l_6$, etc.

Notice that here, $k$ determines the \emph{degree of freedom} of each facility, with larger $k$, the search space for each facility gets larger, while $m$ determines the \emph{breadth} of the search, with larger $m$, we can update the allocation of more facilities simultaneously. In each iteration, we search for a better solution from $(k!)^m$ possible permutations.

Suppose in an iteration, the initial permutation is $\pi: \mathcal{F} \to \mathcal{L}$. Let $\mathcal{F}'=\{F_1, \dots, F_m\}$ denote a collection of $m$ pairwise disjoint refinement subsets, where $F_i \subset \mathcal{F}$ and $F_i \cap F_j = \emptyset$ for $i \not= j$. Since we only allow permutations inside each subset, the QUBO we formulate is equivalent to the formulation (\ref{eq:quboform}) with multiple variables fixed as zero. Namely, $x_{ik} = 0$ if and only if $i \in F_j$ and $k \not\in \{\pi(i) | i \in F_j\}$. with the number of binary variables in (\ref{eq:quboform}) will be reduced to $mk^2$. However, in this formulation, we still have the 2-way 1-hot encoding penalty terms, and therefore is also a C-QUBO-LS.

\begin{figure*}[!htbp]
	\centering
	\includegraphics{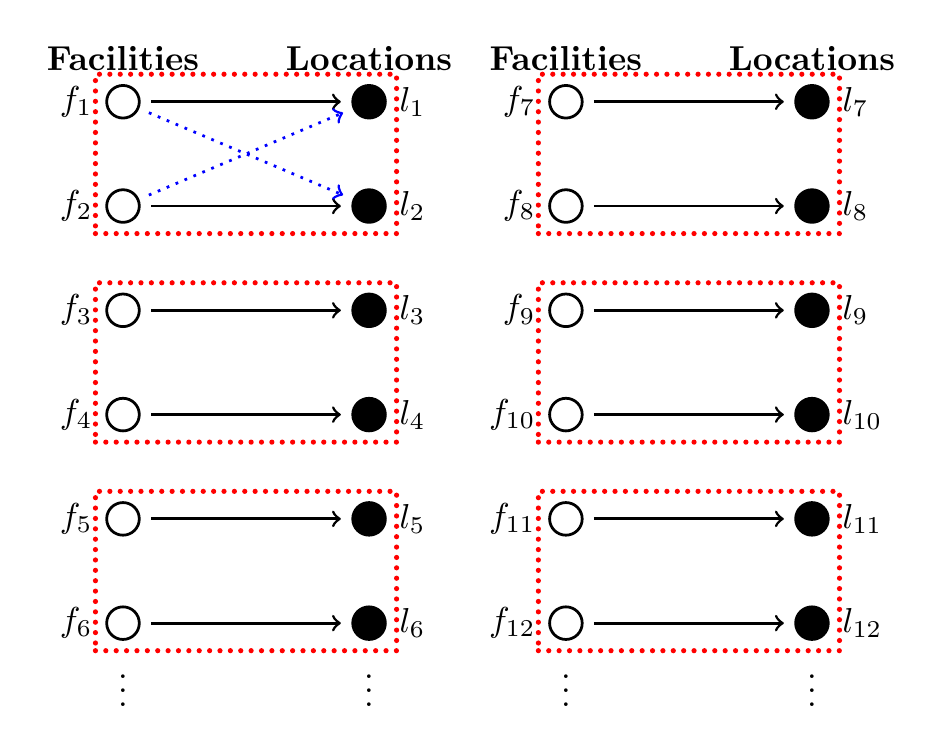}
	\caption{Multiple Local Search Simultaneously with $k=2$}
	\label{fig:multi_ls_k2}
\end{figure*}

Next, we consider the special case when $k = 2$, namely, each subset only contains 2 elements, as shown in Figure \ref{fig:multi_ls_k2}. There are only two possibilities for each subset, swap the current allocation or keep the current allocation. If we formulate this local search with binary variables interpreted as follows:
\begin{equation*}
	x_j = \begin{cases}
		1, & \mbox{if apply pairwise exchange on } F_j, \\
		0, & \mbox{otherwise}.
	\end{cases}
\end{equation*}
We can further reduce the number of variables to $m$, what's more, since the permutation constraints are automatically satisfied after the local change, the 2-way 1-hot encoding in (\ref{eq:quboform}) will be dropped, therefore we can model such local search process into a U-QUBO-LS formulation. Let $\pi$ denote the initial permutation, $\pi_{jj}$ is the permutation that applies pairwise exchange within $F_j$ only, and $\pi_{ij}$ is the permutation that applies pairwise exchange within both $F_i$ and $F_j$. Then the U-QUBO-LS formulation of the local search is given as follows:

\begin{multline*}
\quad \quad \quad	\min    \sum_{F_i, F_j \in \mathcal{F}'}\bigg( \quad x_i\comp{x_j} \sum_{a, b\in F_i\cup F_j} w_{ab}d_{\pi_{ii}(a)\pi_{ii}(b)} \\
	+ \comp{x_i}x_j \sum_{a, b\in F_i\cup F_j} w_{ab}d_{\pi_{jj}(a)\pi_{jj}(b)} \\
 	+ x_ix_j \sum_{a, b\in F_i\cup F_j} w_{ab}d_{\pi_{ij}(a)\pi_{ij}(b)} \\
 	+ \comp{x_i}\comp{x_j} \sum_{a, b\in F_i\cup F_j} w_{ab}d_{\pi(a)\pi(b)} \quad \bigg) \\
 	+ \sum_{F_i\in\mathcal{F}', f \not\in\cup_j F_j} \bigg(\quad x_i \sum_{a, b\in F_i\cup \{f\}} w_{ab}d_{\pi_{ii}(a)\pi_{ii}(b)} \quad\quad\quad \\
 	+ \comp{x_i} \sum_{a, b\in F_i\cup \{f\}} w_{ab}d_{\pi(a)\pi(b)} \quad \bigg). \quad\quad\quad\qquad\qquad
 \end{multline*}

The special case when $k=2$ we present here restrict each local change as pairwise exchange, this can be generalized further, we can define various kinds of local changes, and interpret the binary variables as decision variables: apply this local change or not. With this modeling of local search, the 2-way 1-hot encoding will always be automatically satisfied, meaning no penalizing parameters to scale in the QUBO. Moreover, we use less variables in the formulation.

Another important factor of the algorithm is the order in which the neighborhood is scanned, or in other words, the rules of selecting the subsets. This order can be deterministic or chosen at random. In our experiments, we explore a simple greedy selection rule. Namely, we scan the pair exchange neighborhood, that is, all permutations which can be obtained from the given one by applying a transposition to it. We then rank the pairs by the improvement of the solution, and finally select the subsets based on this ranking.

\subsection{Minimum 2-Sum Problem}
\label{sec:modeling5}

We further demonstrate that U-QUBO-LS can be easily generalized to other combinatorial optimization problems. We use the minimum 2-sum problem (M2sP)  as an example. M2sP is closely related to the problem of reordering a sparse symmetric matrix to reduce its envelope size. The problem is NP-hard for which several effective heuristics have been proposed (see references in \cite{safro2006multilevel}).

Formally, let $G = (V, E)$ be an undirected graph, where $V = \{1, 2, \cdots, n\}$, and $w_{ij} \geq 0, ij \in E$ be the $ij$ edge weight. The goal of M2sP is to find a permutation $\pi: V \to V$ such that the cost
\begin{equation}
\sigma_2(G, \pi) = \sum_{ij\in E} w_{ij}(\pi(i) - \pi(j))^2
\end{equation}
is minimized. This problem can be formulated as QAP \cite{george1997analysis}, and therefore can be formulated as a QUBO shown in (\ref{eq:quboform}) with $w_{ij} = L_{ij}$, where $L$ is the Laplacian matrix of graph $G$, and $d_{ij} = i \cdot j, i, j \in [1, n]$. We then can apply the U-QUBO-LS described in Section \ref{sec:modeling4} to solve the problem.

In our experiments, we use the solution found by spectral algorithm \cite{barnard1995spectral} as the initial solution, which is known to be able to find a relatively good solution for the problem \cite{george1997analysis}, and then use  U-QUBO-LS to iteratively improve the solution.

\subsection{Modeling Local Search Heuristics}
\label{sec:modeling6}

Here we summarize QUBO-LS while comparing our approach to previous methods. Given a combinatorial optimization (CO) problem $P$ with $n$ variables $\{x_i | x_i \in D_i\}_{i=1}^n$, where $D_i$ is the domain of each variable. Let $C$ be the set of constraints among variables, and let $S$ be the set of all feasible solutions of $P$, and the goal of the CO problem is to minimize/maximize an objective function $f(x), x \in S$. Then, the QLS algorithm in \cite{shaydulin2019hybrid,shaydulin2018community,shaydulin2019network} is to first choose a subset $A$ from the $n$ variables, and then search through all possible configurations of $A$ by formulating this sub-problem as a QUBO (see pseudo-code shown in Algorithm \ref{alg:qls}). Note that typically, the QUBO formulations in each iteration of QLS contain penalty terms to model the domain constraints $\{x_i | x_i \in D_i\}_{i=1}^n$ and the variable constraints $C$, thus QLS is a special case of C-QUBO-LS.

\begin{algorithm}
\begin{algorithmic}[1]
\caption{QLS algorithm}
\label{alg:qls}
\STATE \textbf{Input:} Problem $P$
\STATE Initialize a feasible solution $x \in S$ of problem $P$ of size $n$
\FOR {$i = 1, 2, \cdots$}
\STATE Choose a subset of the $n$ variables $A \subset \{x_i\}_{i=1}^n$
\STATE Reconfigure $A$ by formulating it into a QUBO
\STATE Use a QUBO solver to find a new feasible solution
\IF{new solution is better}
\STATE accept the new solution
\ENDIF
\ENDFOR
\end{algorithmic}
\end{algorithm}

We extend QLS in such a way that instead of only choosing one subset from the $n$ possible variables, we choose $m$ pairwise disjoint subsets from the $n$ variables. Within each subset, we search through all possible feasible configurations. The pseudo-code of C-QUBO-LS is given in Algorithm \ref{alg:general}. In each iteration of Algorithm \ref{alg:general} we create an quadratic binary optimization problem that has constraints, therefore there is the need to introduce penalty terms to the QUBO. However, if we map a binary variable to the decision of whether or not a specified local change should be applied, by the nature of these local changes, the constraints will not be violated and thus no penalty terms in the QUBO formulation (see psedo-code of U-QUBO-LS in Algorithm \ref{alg:noconstraint}). Here, this local change can be the swapping based local change that we discussed in the previous sections, it can also be other types of local changes: cyclic exchange, recombination operator in genetic algorithms etc.

\begin{algorithm}
\begin{algorithmic}[1]
\caption{C-QUBO-LS}
\label{alg:general}
\STATE \textbf{Input:} Problem $P$
\STATE Initialize a feasible solution $x \in S$ of problem $P$ of size $n$
\FOR {$i = 1, 2, \cdots$}
\STATE Choose $m$ pairwise disjoint subsets of the $n$ variables $A_1, \dots, A_m \subset \{x_i\}_{i=1}^n$, such that $A_j \cap A_k = \text{\O}, j \neq k$
\STATE Reconfigure $\{A_j\}_{j=1}^m$ by formulating it into a QUBO with penalty terms to model the constraints
\STATE Use a QUBO solver to find a new feasible solution
\IF{new solution is better}
\STATE accept the new solution
\ENDIF
\ENDFOR
\end{algorithmic}
\end{algorithm}

\begin{algorithm}
\begin{algorithmic}[1]
\STATE Initialize a feasible solution $x \in S$ of problem $P$ of size $n$
\FOR {$i = 1, 2, \cdots$}
\STATE Choose $m$ pairwise disjoint subsets of the $n$ variables $A_1, \dots, A_m \subset \{x_i\}_{i=1}^n$, such that $A_j \cap A_k = \text{\O}, j \neq k$
\STATE Define a possible local change for each subset $A_j$ and assign a binary variable to represent whether or not to apply that local change
\STATE Formulate the unconstrained QUBO and use QUBO solver to find a solution
\IF{new solution is better}
\STATE Accept the new solution
\ENDIF
\ENDFOR
\end{algorithmic}
\caption{U-QUBO-LS}
\label{alg:noconstraint}
\end{algorithm}

With our approach, for problems with 2-way 1-hot constraints like TSP and QAP, in each iteration, suppose we select $m$ subsets of size $k$, then we solve a QUBO to explore the search space of size $(k!)^m$. With $k=2$, we will utilize the hardware more efficiently since we can model it into a U-QUBO-LS, and all solutions searched by the QUBO solver are feasible solutions, while for other values of $k$, a large number of the solutions searched are infeasible since they will violate the 2-way 1-hot constraints.

Finally, in general, parallel local search requires that if more than one local search is performed at the same iteration, these sub-problems must be mutually independent, namely, the local changes applied in parallel should return the same result as if they are applied in sequence. Whereas our approach does not have this requirement making it a more powerful approach.

\section{Computational Results}
\label{sec:experiment}

We demonstrate the performance of QUBO-LS with  QAP and M2sP. We choose QAP mainly because a large class of problems form special cases of QAP. The main goal of the experiments is to \emph{compare the different modeling strategies} for local search on special-purpose hardware of fixed size. To demonstrate the efficacy of our approach, we also compare it with simulated annealing (SA) \cite{kirkpatrick1983optimization}.

All algorithms are implemented in Python 3.7. For IPU, we use the Fujitsu Digital Annealing Unit (DA) with the parallel tempering mode. For each QUBO formulated, we may use up to 1024 binary variables.

\subsection{Experiments on QAP}
\label{sec:experiment1}

We test QUBO-LS on problem instances from the QAP benchmark library \texttt{QAPLIB} \cite{burkard1997qaplib} (\url{http://anjos.mgi.polymtl.ca/qaplib/}). Here we present the results of instances of size larger than 100.

\begin{figure*}[!htbp]
	\centering
	\includegraphics[width=\linewidth]{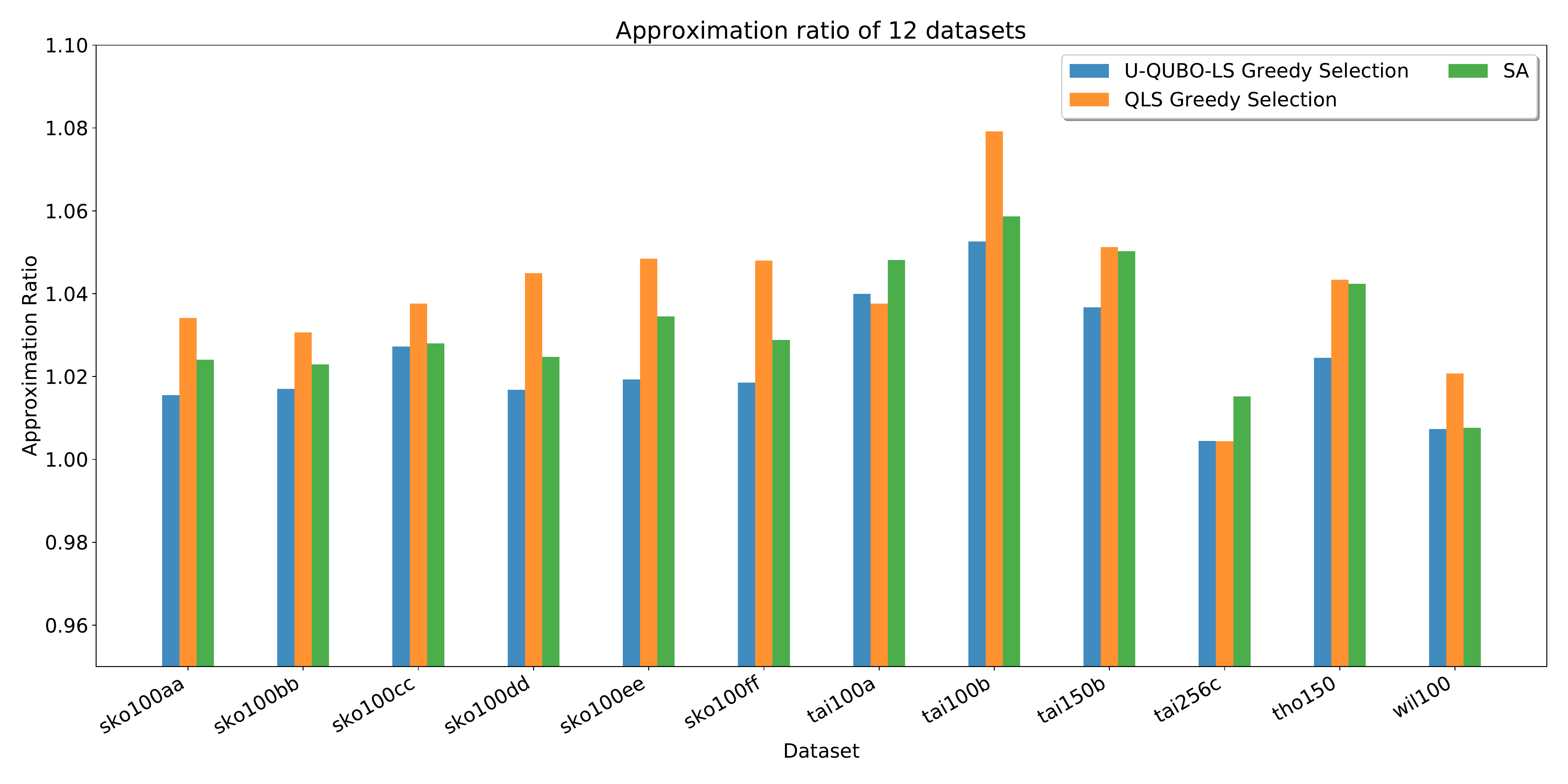}
	\caption{Computational results of 12 QAP instances, U-QUBO-LS achieves best result in 10 of them.}
	\label{fig:qap_result}
\end{figure*}

We first compare the performance of U-QUBO-LS, the QLS Algorithm \ref{alg:qls} \cite{shaydulin2019hybrid,shaydulin2018community,shaydulin2019network},  and SA \cite{kirkpatrick1983optimization} in Figure \ref{fig:qap_result}. For QLS and U-QUBO-LS, we test with a simple selection strategy: greedy selection. That is, we scan the pair exchange neighborhood and then rank the pairs by the improvement of the solution, and finally select the subsets based on this ranking. We allow at most 30 iterations for both algorithms. For SA, we allow at most 10000 iterations. For all instances, we start with a randomly generated permutation as the initial solution, and all algorithms start from the same initial solution. The approximation ratio is the ratio with numerator equal to the objective value we obtained upon termination of each algorithm and the denominator equal to the best known value. We can see that in general, U-QUBO-LS performs better than QLS and SA in terms of solution quality.

Next, we further look at the performance of  C-QUBO-LS and U-QUBO-LS in more detail. There are mainly two parameters in DA we tuned in our experiments. The first is the number of  MC steps in the DA algorithm \cite{aramon2019physics} when solving a QUBO. As pointed out in \cite{aramon2019physics}, each MC step takes the same amount of time, therefore can be considered as a time limit on DA. The second is the choice of initial  binary solution of each QUBO. We carried out experiments in both setting, (i) initializing from a random binary solution and (ii) initializing from a given binary solution.

\begin{figure*}
\hfill
\subfigure[random initial binary solution in DA]{\includegraphics{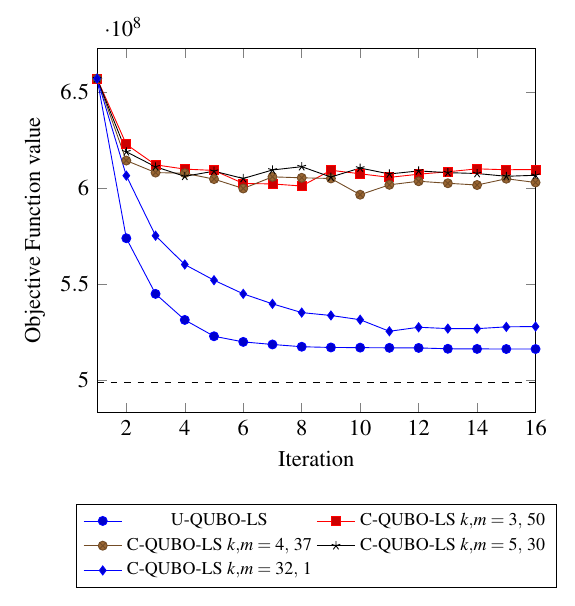}}
\hfill
\subfigure[assign initial binary solution to DA]{\includegraphics{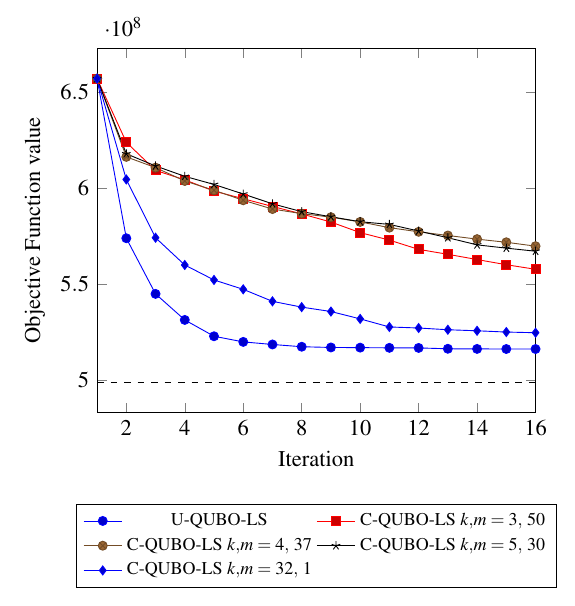}}
\hfill
\caption{Comparison of U-QUBO-LS and C-QUBO-LS, greedy selection. Dataset: \texttt{tai150b}. U-QUBO-LS reach close to the best known solution (gap of 4.21$\%$) after 5 iterations.}
\label{fig:comparekm1}
\end{figure*}

\begin{figure*}
\hfill
\subfigure[random initial binary solution in DA]{\includegraphics{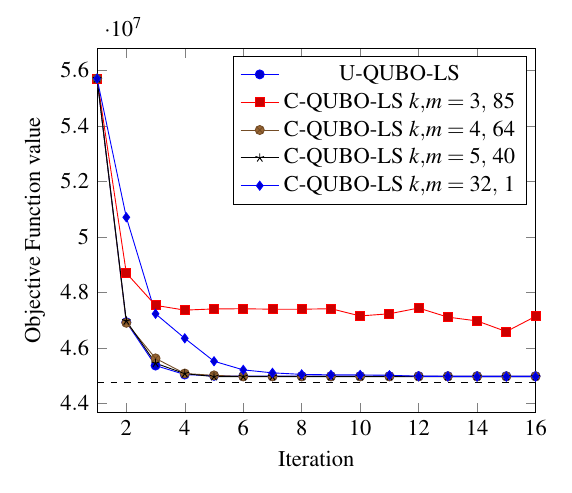}}
\hfill
\subfigure[assign initial binary solution to DA]{\includegraphics{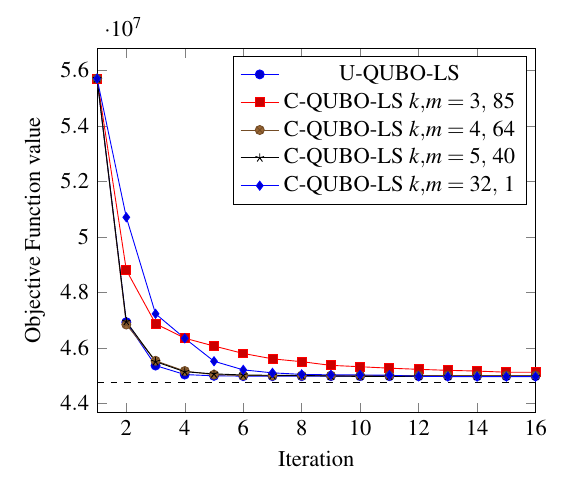}}
\hfill
\caption{Comparison of U-QUBO-LS and C-QUBO-LS, greedy selection. Dataset: \texttt{tai256c}. U-QUBO-LS reach close to the best known solution (gap of 0.49$\%$) after 5 iterations.}
\label{fig:comparekm2}
\end{figure*}

\begin{figure*}
\hfill
\subfigure[random initial binary solution in DA]{\includegraphics{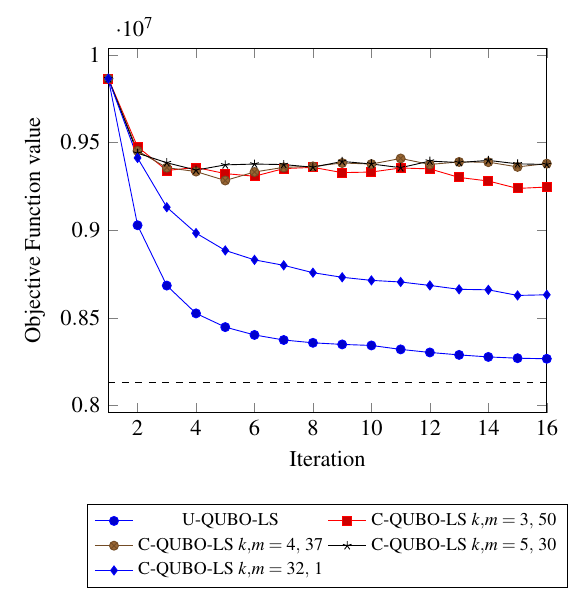}}
\hfill
\subfigure[assign initial binary solution to DA]{\includegraphics{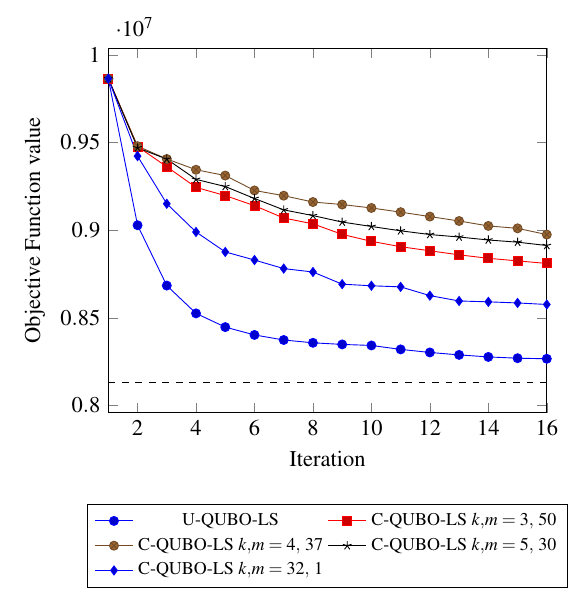}}
\hfill
\caption{Comparison of U-QUBO-LS and C-QUBO-LS, greedy selection. Dataset: \texttt{tho150}. U-QUBO-LS reach close to the best known solution (gap of 3.31$\%$) after 5 iterations.}
\label{fig:comparekm3}
\end{figure*}

We compare the performance of U-QUBO-LS and C-QUBO-LS with different values of $k$ and $m$. Note that in each iteration, for a QAP instance with $n$ facilities, with U-QUBO-LS, we are solving a QUBO with only $\min (n/2, 1024)$ binary variables, and for C-QUBO-LS, we are solving a QUBO with $\min(k^2m, 1024)$ variables. Figure \ref{fig:comparekm1} gives the value of the objective function with respect to the number of iterations of the algorithm with dataset \texttt{tai150b}. In Figure \ref{fig:comparekm1} (a), we start with a randomly generated permutation, and apply greedy selection rule to select the refinement subsets. With U-QUBO-LS, we can reach close to the best known solution (gap of 4.21$\%$) after 5 iterations, and the quality of the solution is the best among all  algorithms. Figure \ref{fig:comparekm2} (a) and \ref{fig:comparekm3} (a) gives results of dataset \texttt{tai256c} and \texttt{tho150}, with  U-QUBO-LS, after 5 iterations, the gap between the solution we found and the best known solution is 0.49$\%$ and 3.31$\%$ respectively.

Since at each iteration, we are solving a QUBO in the DA, the configuration in DA will also effect the quality of the solution. Figure \ref{fig:comparekm1}-\ref{fig:comparekm3} (a) show the results when we did not specify an initial binary solution to DA in the annealing process, namely, DA will randomly generate a binary string as initial solution. Figure \ref{fig:comparekm1}-\ref{fig:comparekm3} (b) show the results when we specify the binary encoding of the current permutation as the initial binary solution to DA. The advantage of specifying an initial binary solution is that we will never get a solution that is worse than the initial feasible solution given, and the drawback is that we will be more likely to stay in a local optimal. From the plots, we can see that U-QUBO-LS in both cases achieve the best result among all algorithms.

\begin{figure*}
\hfill
\subfigure[C-QUBO-LS $k, m = 32, 1$]{\includegraphics{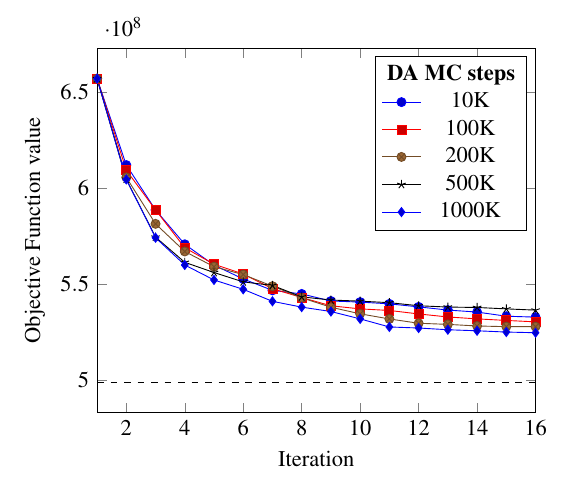}}
\hfill
\subfigure[U-QUBO-LS]{\includegraphics{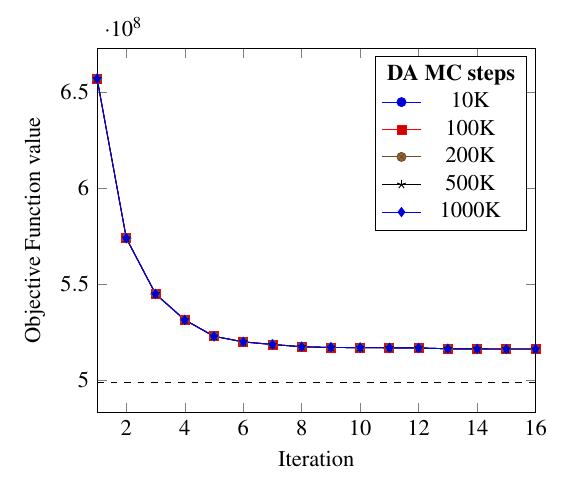}}
\hfill
\caption{Comparison of U-QUBO-LS and C-QUBO-LS, greedy selection, assign initial binary solution to DA. Dataset: \texttt{tai150b}. C-QUBO-LS needs more MC steps (time per iteration) comparing to U-QUBO-LS.}
\label{fig:compareiter1}
\end{figure*}

\begin{figure*}
\hfill
\subfigure[Dataset: \texttt{tai256c}]{\includegraphics{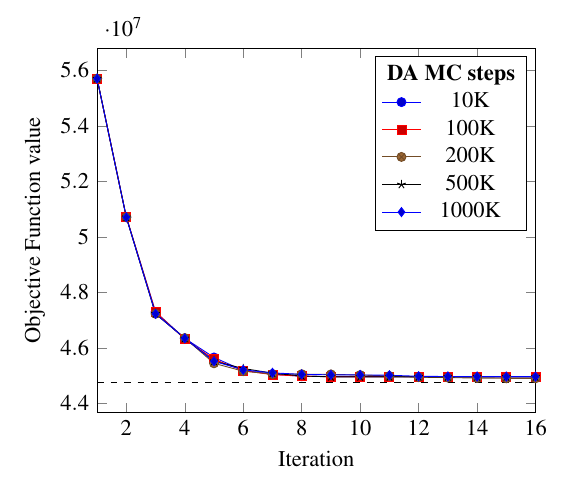}}
\hfill
\subfigure[Dataset: \texttt{tho150}]{\includegraphics{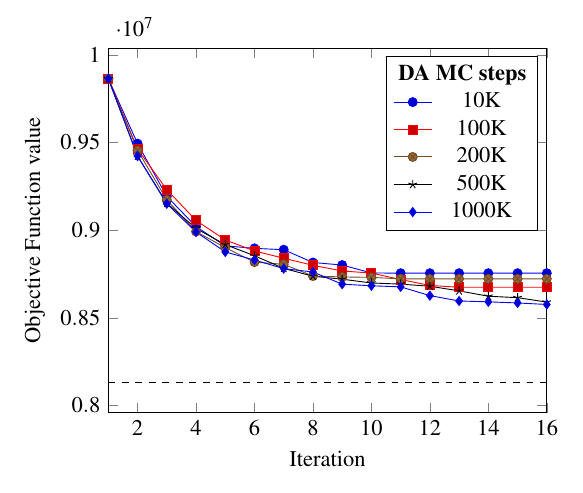}}
\hfill
\caption{Dependence of objective value on the number of iterations using the greedy selection, assign initial binary solution to DA, C-QUBO-LS $k, m = 32, 1$.}
\label{fig:compareiter2}
\end{figure*}

\paragraph{Time limit per iteration:}
Next, we compare the performance of the modeling strategies with different time limit for solving the QUBO. We give the DA different number of MC steps performed, to control the annealing time. As shown in Figure \ref{fig:compareiter1}, we found that with U-QUBO-LS, with a small number of  MC steps (10,000), we can find a solution that has good quality, and we observe the same for all other instances. While for C-QUBO-LS, in most cases, as shown in Figure \ref{fig:compareiter2}, we need more MC steps in DA to guarantee the quality of the solution.

\subsection{Experiments on M2sP}
\label{sec:experiment2}

We also test QUBO-LS on M2sP as a post-processing refinement method. That is, instead of starting with a random initial solution, we start with the solution obtained by the spectral ordering \cite{barnard1995spectral}, which is known to be able to find a relatively good solution for the M2sP problem. All algorithms start from the same initial solution. We test on problem instances generated by \texttt{Networkx} and instances from Benchmark Graphs for Practical Graph Isomorphism \cite{neuen2017benchmark}  (\url{https://www.lics.rwth-aachen.de/go/id/rtok/}).

\begin{figure*}[!htbp]
	\centering
	\includegraphics[width=\linewidth]{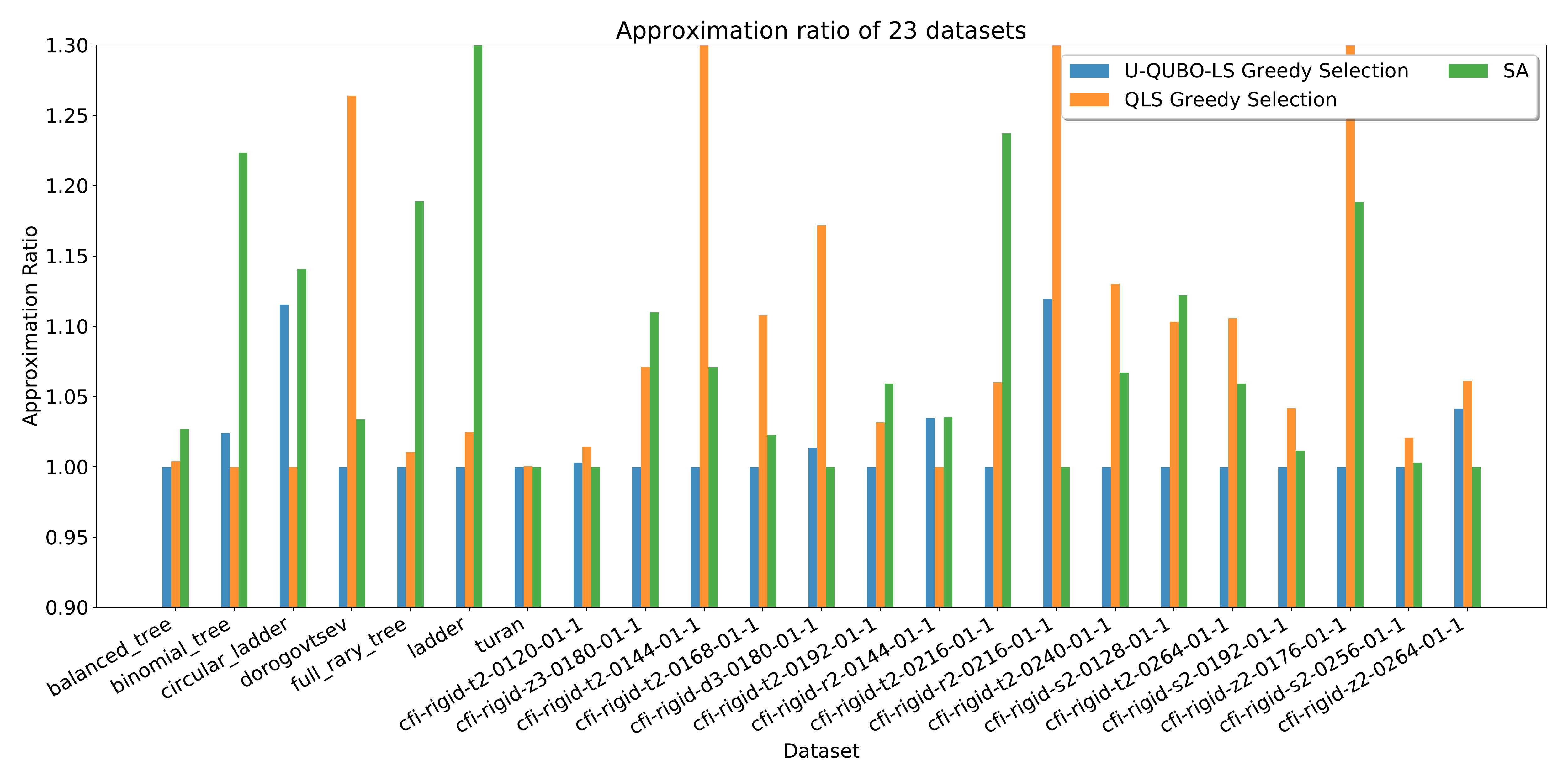}
	\caption{Computational results of 23 M2sP instances, U-QUBO-LS achieves best result in 16 of them.}\label{fig:min2sum}
\end{figure*}

We compare the performance of U-QUBO-LS, QLS \cite{shaydulin2019hybrid,shaydulin2018community,shaydulin2019network}, and SA \cite{kirkpatrick1983optimization} in Figure \ref{fig:min2sum} and provide more details in Table \ref{tab:min2sum_result}. Similar to the QAP experiments, for QLS and U-QUBO-LS, we test with a simple selection strategy: greedy selection. That is, we scan the pair exchange neighborhood and then rank the pairs by the improvement of the solution, and finally select the subsets based on this ranking. We allow at most 30 iterations for both algorithm. And for SA, we allow at most 15000 iterations. We can see from table \ref{tab:min2sum_result}, again, U-QUBO-LS achieves a better solution most often. Note that in each iteration, for a problem instance with $|V| = n$ vertices, with U-QUBO-LS, we are solving a QUBO with only $\min (n/2, 1024)$ binary variables, and for QLS, we are solving a QUBO with 1024 variables.

\begin{table}
\centering
\caption{Computational results of 23 M2sP instances, U-QUBO-LS achieves best result in 16 of them.}
\label{tab:min2sum_result}
\begin{tabular}{c|c|c|c|c|c}
problem & $|V|$ & $|E|$ & U-QUBO-LS & QLS & SA\\
\hline
balanced tree & 156 & 155 & \boldblue{10266} & 10306 & 10541 \\
binomial tree & 128 & 127 & 2892 & \boldblue{2824} & 3455 \\
circular ladder & 120 & 180 & 12422 & \boldblue{11136} & 12702 \\
dorogovtsev & 123 & 243 & \boldblue{53856} & 68082 & 55672 \\
full rary tree & 120 & 119 & \boldblue{4572} & 4620 & 5436 \\
ladder & 110 & 163 & \boldblue{487} & 499 & 3771 \\
turan & 110 & 4537 & \boldblue{9107533} & 9109708 & 9107588 \\
cfi-rigid-d3-0180-01-1 & 180 & 864 & 1136968 & 1314408 & \boldblue{1121720} \\
cfi-rigid-t2-0120-01-1 & 120 & 964 & 593188 & 599954 & \boldblue{591452} \\
cfi-rigid-t2-0144-01-1 & 144 & 1148 & \boldblue{866406} & 1148630 & 927785 \\
cfi-rigid-t2-0168-01-1 & 168 & 1340 & \boldblue{1253632} & 1388563 & 1282149 \\
cfi-rigid-t2-0192-01-1 & 192 & 1548 & \boldblue{1945027} & 2006573 & 2060212 \\
cfi-rigid-t2-0216-01-1 & 216 & 1748 & \boldblue{2394173} & 2538142 & 2962434 \\
cfi-rigid-t2-0240-01-1 & 240 & 1956 & \boldblue{2886686} & 3262181 & 3080522 \\
cfi-rigid-t2-0264-01-1 & 264 & 2152 & \boldblue{5032800} & 5564638 & 5330893 \\
cfi-rigid-r2-0144-01-1 & 144 & 288 & 127766 & \boldblue{123468} & 127828 \\
cfi-rigid-r2-0216-01-1 & 216 & 432 & 321235 & 858697 & \boldblue{286978} \\
cfi-rigid-s2-0128-01-1 & 128 & 1312 & \boldblue{1372005} & 1513711 & 1539419 \\
cfi-rigid-s2-0192-01-1 & 192 & 1984 & \boldblue{3674022} & 3826696 & 3716500 \\
cfi-rigid-s2-0256-01-1 & 256 & 2656 & \boldblue{7914054} & 8076603 & 7938319 \\
cfi-rigid-z2-0176-01-1 & 176 & 384 & \boldblue{290583} & 382316 & 345359 \\
cfi-rigid-z2-0264-01-1 & 264 & 576 & 1088787 & 1109326 & \boldblue{1045444} \\
cfi-rigid-z3-0180-01-1 & 180 & 432 & \boldblue{553701} & 593020 & 614510 \\
\end{tabular}
\end{table}

\section{Conclusion}
\label{sec:conclusion}
Post Moore systems such as special-purpose hardware are being developed for different scientific domains. In combinatorial optimization, several novel hardware types are  emerging with a common purpose, that is, solving the Ising model (or QUBO). As the hardware emerges, there is a challenge for existing well-established algorithms to take advantage of these systems. This is especially important when the problem is  large. We have tackled this challenge by proposing a QUBO-LS framework including different models, modeling techniques and algorithms that utilize the hardware efficiently. We classify QUBO-LS into two categories: C-QUBO-LS and U-QUBO-LS. In particular, for large problems, we have demonstrated how to model multiple sub-problems that are not necessarily mutually independent as a step in a local search framework. Given that the QUBO is unconstrained by definition, we have further showed how to model sub-problems as a U-QUBO-LS that implicitly satisfy the constraints, thus searching an exponentially larger search space per iteration compared to previous methods and also utilizing the given hardware more efficiently. This provides new possibilities to escape from local optima. Our novel modeling techniques and algorithms can be easily adopted to a large class of local search algorithms.

\section*{Conflict of interest}
The authors declare that they have no conflict of interest.

\bibliographystyle{spmpsci}      
\bibliography{bib}   

\begin{thebibliography}{10}
\providecommand{\url}[1]{{#1}}
\providecommand{\urlprefix}{URL }
\expandafter\ifx\csname urlstyle\endcsname\relax
  \providecommand{\doi}[1]{DOI~\discretionary{}{}{}#1}\else
  \providecommand{\doi}{DOI~\discretionary{}{}{}\begingroup
  \urlstyle{rm}\Url}\fi

\bibitem{aramon2019physics}
Aramon, M., Rosenberg, G., Valiante, E., Miyazawa, T., Tamura, H., Katzgrabeer,
  H.: Physics-inspired optimization for quadratic unconstrained problems using
  a digital annealer.
\newblock Frontiers in Physics \textbf{7}, 48 (2019)

\bibitem{barnard1995spectral}
Barnard, S.T., Pothen, A., Simon, H.: A spectral algorithm for envelope
  reduction of sparse matrices.
\newblock Numerical linear algebra with applications \textbf{2}(4), 317--334
  (1995)

\bibitem{bock1958algorithm}
Bock, F.: An algorithm for solving travelling-salesman and related network
  optimization problems.
\newblock In: Operations Research, vol.~6, pp. 897--897. INST OPERATIONS
  RESEARCH MANAGEMENT SCIENCES 901 ELKRIDGE LANDING RD, STE~… (1958)

\bibitem{boman2012zoltan}
Boman, E.G., {\c{C}}ataly{\"u}rek, {\"U}.V., Chevalier, C., Devine, K.D.: The
  zoltan and isorropia parallel toolkits for combinatorial scientific
  computing: Partitioning, ordering and coloring.
\newblock Scientific Programming \textbf{20}(2), 129--150 (2012)

\bibitem{booth2017partitioning}
Booth, M., Reinhardt, S.P., Roy, A.: Partitioning optimization problems for
  hybrid classical.
\newblock quantum execution. Technical Report pp. 01--09 (2017)

\bibitem{bulucc2016recent}
Bulu{\c{c}}, A., Meyerhenke, H., Safro, I., Sanders, P., Schulz, C.: Recent
  advances in graph partitioning.
\newblock In: Algorithm Engineering, pp. 117--158. Springer (2016)

\bibitem{burkard1998quadratic}
Burkard, R.E., Cela, E., Pardalos, P.M., Pitsoulis, L.S.: The quadratic
  assignment problem.
\newblock In: Handbook of combinatorial optimization, pp. 1713--1809. Springer
  (1998)

\bibitem{burkard1997qaplib}
Burkard, R.E., Karisch, S.E., Rendl, F.: Qaplib--a quadratic assignment problem
  library.
\newblock Journal of Global optimization \textbf{10}(4), 391--403 (1997)

\bibitem{coffrin2019evaluating}
Coffrin, C., Nagarajan, H., Bent, R.: Evaluating ising processing units with
  integer programming.
\newblock In: International Conference on Integration of Constraint
  Programming, Artificial Intelligence, and Operations Research, pp. 163--181.
  Springer (2019)

\bibitem{crawford2016reinforcement}
Crawford, D., Levit, A., Ghadermarzy, N., Oberoi, J.S., Ronagh, P.:
  Reinforcement learning using quantum boltzmann machines.
\newblock arXiv preprint arXiv:1612.05695  (2016)

\bibitem{dash2013note}
Dash, S.: A note on qubo instances defined on chimera graphs.
\newblock arXiv preprint arXiv:1306.1202  (2013)

\bibitem{farhi2000quantum}
Farhi, E., Goldstone, J., Gutmann, S., Sipser, M.: Quantum computation by
  adiabatic evolution.
\newblock arXiv preprint quant-ph/0001106  (2000)

\bibitem{fiduccia1982linear}
Fiduccia, C.M., Mattheyses, R.M.: A linear-time heuristic for improving network
  partitions.
\newblock In: 19th Design Automation Conference, pp. 175--181. IEEE (1982)

\bibitem{george1997analysis}
George, A., Pothen, A.: An analysis of spectral envelope reduction via
  quadratic assignment problems.
\newblock SIAM Journal on Matrix Analysis and Applications \textbf{18}(3),
  706--732 (1997)

\bibitem{glover2018tutorial}
Glover, F., Kochenberger, G.: A tutorial on formulating qubo models.
\newblock arXiv preprint arXiv:1811.11538  (2018)

\bibitem{glover1998adaptive}
Glover, F., Kochenberger, G.A., Alidaee, B.: Adaptive memory tabu search for
  binary quadratic programs.
\newblock Management Science \textbf{44}(3), 336--345 (1998)

\bibitem{hamze2004fields}
Hamze, F., de~Freitas, N.: From fields to trees.
\newblock In: Proceedings of the 20th conference on Uncertainty in artificial
  intelligence, pp. 243--250. AUAI Press (2004)

\bibitem{hastings1970monte}
Hastings, W.K.: Monte carlo sampling methods using markov chains and their
  applications  (1970)

\bibitem{henderson2018leveraging}
Henderson, M., Novak, J., Cook, T.: Leveraging adiabatic quantum computation
  for election forecasting.
\newblock arXiv preprint arXiv:1802.00069  (2018)

\bibitem{hernandez2017enhancing}
Hernandez, M., Aramon, M.: Enhancing quantum annealing performance for the
  molecular similarity problem.
\newblock Quantum Information Processing \textbf{16}(5), 133 (2017)

\bibitem{hernandez2016novel}
Hernandez, M., Zaribafiyan, A., Aramon, M., Naghibi, M.: A novel graph-based
  approach for determining molecular similarity.
\newblock arXiv preprint arXiv:1601.06693  (2016)

\bibitem{inagaki2016coherent}
Inagaki, T., Haribara, Y., Igarashi, K., Sonobe, T., Tamate, S., Honjo, T.,
  Marandi, A., McMahon, P.L., Umeki, T., Enbutsu, K., et~al.: A coherent ising
  machine for 2000-node optimization problems.
\newblock Science \textbf{354}(6312), 603--606 (2016)

\bibitem{johnson2011quantum}
Johnson, M.W., Amin, M.H., Gildert, S., Lanting, T., Hamze, F., Dickson, N.,
  Harris, R., Berkley, A.J., Johansson, J., Bunyk, P., et~al.: Quantum
  annealing with manufactured spins.
\newblock Nature \textbf{473}(7346), 194 (2011)

\bibitem{kadowaki1998quantum}
Kadowaki, T., Nishimori, H.: Quantum annealing in the transverse ising model.
\newblock Physical Review E \textbf{58}(5), 5355 (1998)

\bibitem{karypis1998fast}
Karypis, G., Kumar, V.: A fast and high quality multilevel scheme for
  partitioning irregular graphs.
\newblock SIAM Journal on scientific Computing \textbf{20}(1), 359--392 (1998)

\bibitem{kernighan1970efficient}
Kernighan, B.W., Lin, S.: An efficient heuristic procedure for partitioning
  graphs.
\newblock Bell system technical journal \textbf{49}(2), 291--307 (1970)

\bibitem{khoshaman2018quantum}
Khoshaman, A., Vinci, W., Denis, B., Andriyash, E., Amin, M.H.: Quantum
  variational autoencoder.
\newblock Quantum Science and Technology \textbf{4}(1), 014001 (2018)

\bibitem{kielpinski2016information}
Kielpinski, D., Bose, R., Pelc, J., Van~Vaerenbergh, T., Mendoza, G., Tezak,
  N., Beausoleil, R.G.: Information processing with large-scale optical
  integrated circuits.
\newblock In: 2016 IEEE International Conference on Rebooting Computing (ICRC),
  pp. 1--4. IEEE (2016)

\bibitem{kirkpatrick1983optimization}
Kirkpatrick, S., Gelatt, C.D., Vecchi, M.P.: Optimization by simulated
  annealing.
\newblock science \textbf{220}(4598), 671--680 (1983)

\bibitem{kochenberger2006unified}
Kochenberger, G.A., Glover, F.: A unified framework for modeling and solving
  combinatorial optimization problems: A tutorial.
\newblock In: Multiscale Optimization Methods and Applications, pp. 101--124.
  Springer (2006)

\bibitem{levit2017free}
Levit, A., Crawford, D., Ghadermarzy, N., Oberoi, J.S., Zahedinejad, E.,
  Ronagh, P.: Free energy-based reinforcement learning using a quantum
  processor.
\newblock arXiv preprint arXiv:1706.00074  (2017)

\bibitem{lucas2014ising}
Lucas, A.: Ising formulations of many np problems.
\newblock Frontiers in Physics \textbf{2}, 5 (2014)

\bibitem{mcgeoch2013experimental}
McGeoch, C.C., Wang, C.: Experimental evaluation of an adiabiatic quantum
  system for combinatorial optimization.
\newblock In: Proceedings of the ACM International Conference on Computing
  Frontiers, p.~23. ACM (2013)

\bibitem{mcmahon2016fully}
McMahon, P.L., Marandi, A., Haribara, Y., Hamerly, R., Langrock, C., Tamate,
  S., Inagaki, T., Takesue, H., Utsunomiya, S., Aihara, K., et~al.: A fully
  programmable 100-spin coherent ising machine with all-to-all connections.
\newblock Science \textbf{354}(6312), 614--617 (2016)

\bibitem{metropolis1953equation}
Metropolis, N., Rosenbluth, A.W., Rosenbluth, M.N., Teller, A.H., Teller, E.:
  Equation of state calculations by fast computing machines.
\newblock The journal of chemical physics \textbf{21}(6), 1087--1092 (1953)

\bibitem{negre2019detecting}
Negre, C.F., Ushijima-Mwesigwa, H., Mniszewski, S.M.: Detecting multiple
  communities using quantum annealing on the d-wave system.
\newblock arXiv preprint arXiv:1901.09756  (2019)

\bibitem{neuen2017benchmark}
Neuen, D., Schweitzer, P.: Benchmark graphs for practical graph isomorphism.
\newblock arXiv preprint arXiv:1705.03686  (2017)

\bibitem{neukart2017traffic}
Neukart, F., Compostella, G., Seidel, C., Von~Dollen, D., Yarkoni, S., Parney,
  B.: Traffic flow optimization using a quantum annealer.
\newblock Frontiers in ICT \textbf{4}, 29 (2017)

\bibitem{nightingale1998quantum}
Nightingale, M.P., Umrigar, C.J.: Quantum Monte Carlo methods in physics and
  chemistry.
\newblock 525. Springer Science \& Business Media (1998)

\bibitem{pisinger2010large}
Pisinger, D., Ropke, S.: Large neighborhood search.
\newblock In: Handbook of metaheuristics, pp. 399--419. Springer (2010)

\bibitem{puget2018d}
Puget, J.: D-wave vs cplex comparison. part 2: Qubo, 2013 (2018)

\bibitem{rosenberg2016solving}
Rosenberg, G., Haghnegahdar, P., Goddard, P., Carr, P., Wu, K., De~Prado, M.L.:
  Solving the optimal trading trajectory problem using a quantum annealer.
\newblock IEEE Journal of Selected Topics in Signal Processing \textbf{10}(6),
  1053--1060 (2016)

\bibitem{safro2006multilevel}
Safro, I., Ron, D., Brandt, A.: A multilevel algorithm for the minimum 2-sum
  problem.
\newblock J. Graph Algorithms Appl. \textbf{10}(2), 237--258 (2006)

\bibitem{safro2015advanced}
Safro, I., Sanders, P., Schulz, C.: Advanced coarsening schemes for graph
  partitioning.
\newblock Journal of Experimental Algorithmics (JEA) \textbf{19}, 2--2 (2015)

\bibitem{sanders2011engineering}
Sanders, P., Schulz, C.: Engineering multilevel graph partitioning algorithms.
\newblock In: European Symposium on Algorithms, pp. 469--480. Springer (2011)

\bibitem{schaller1997moore}
Schaller, R.R.: Moore's law: past, present and future.
\newblock IEEE spectrum \textbf{34}(6), 52--59 (1997)

\bibitem{schneidman2006}
Schneidman, E., Berry~II, M., Segev, R., Bialek, W.: Weak pairwise correlations
  imply strongly correlated network states in a neural population.
\newblock Nature \textbf{440}, 1007--12 (2006).
\newblock \doi{10.1038/nature04701}

\bibitem{selby2014efficient}
Selby, A.: Efficient subgraph-based sampling of ising-type models with
  frustration.
\newblock arXiv preprint arXiv:1409.3934  (2014)

\bibitem{shaydulin2019relaxation}
Shaydulin, R., Chen, J., Safro, I.: Relaxation-based coarsening for multilevel
  hypergraph partitioning.
\newblock Multiscale Modeling \& Simulation \textbf{17}(1), 482--506 (2019)

\bibitem{shaydulin2019hybrid}
Shaydulin, R., Ushijima-Mwesigwa, H., Negre, C.F., Safro, I., Mniszewski, S.M.,
  Alexeev, Y.: A hybrid approach for solving optimization problems on small
  quantum computers.
\newblock Computer \textbf{52}(6), 18--26 (2019)

\bibitem{shaydulin2018community}
Shaydulin, R., Ushijima-Mwesigwa, H., Safro, I., Mniszewski, S., Alexeev, Y.:
  Community detection across emerging quantum architectures.
\newblock 3rd International Workshop on Post Moore's Era Supercomputing (PMES
  2018)  (2018)

\bibitem{shaydulin2019network}
Shaydulin, R., Ushijima-Mwesigwa, H., Safro, I., Mniszewski, S., Alexeev, Y.:
  Network community detection on small quantum computers.
\newblock Advanced Quantum Technologies \textbf{2}(9), 1900029 (2019)

\bibitem{swendsen1986replica}
Swendsen, R.H., Wang, J.S.: Replica monte carlo simulation of spin-glasses.
\newblock Physical review letters \textbf{57}(21), 2607 (1986)

\bibitem{terry2019quantum}
Terry, J.P., Akrobotu, P.D., Negre, C.F., Mniszewski, S.M.: Quantum isomer
  search.
\newblock arXiv preprint arXiv:1908.00542  (2019)

\bibitem{ushijima2017graph}
Ushijima-Mwesigwa, H., Negre, C.F., Mniszewski, S.M.: Graph partitioning using
  quantum annealing on the d-wave system.
\newblock In: Proceedings of the Second International Workshop on Post Moores
  Era Supercomputing, pp. 22--29. ACM (2017)

\bibitem{multilevel}
Ushijima-Mwesigwa, H., Shaydulin, R., Negre, C.F., Mniszewski, S.M., Alexeev,
  Y., Safro, I.: Multilevel combinatorial optimization across quantum
  architectures.
\newblock accepted in ACM Transactions on Quantum Computing, preprint at
  arXiv:1910.09985  (2020)

\bibitem{wang2012path}
Wang, Y., L{\"u}, Z., Glover, F., Hao, J.K.: Path relinking for unconstrained
  binary quadratic programming.
\newblock European Journal of Operational Research \textbf{223}(3), 595--604
  (2012)

\bibitem{yamaoka201524}
Yamaoka, M., Yoshimura, C., Hayashi, M., Okuyama, T., Aoki, H., Mizuno, H.:
  24.3 20k-spin ising chip for combinational optimization problem with cmos
  annealing.
\newblock In: 2015 IEEE International Solid-State Circuits Conference-(ISSCC)
  Digest of Technical Papers, pp. 1--3. IEEE (2015)

\bibitem{yoshimura2013spatial}
Yoshimura, C., Yamaoka, M., Aoki, H., Mizuno, H.: Spatial computing
  architecture using randomness of memory cell stability under voltage control.
\newblock In: 2013 European Conference on Circuit Theory and Design (ECCTD),
  pp. 1--4. IEEE (2013)

\end{thebibliography}


\end{document}